# INTERNATIONAL DATA CENTRE: REVIEWED EVENT BULLETIN VS. WAVEFORM CROSS-CORELATION BULLETIN

Dmitry Bobrov, John Coyne, Jeffrey Given, Urtnasan Khukhuudei, Ivan Kitov, Kirill Sitnikov, Spilio Spiliopoulos, and Lassina Zerbo

Comprehensive Nuclear-Test-Ban Treaty Organization

**Abstract**
Our objective is to assess the performance of waveform cross-correlation technique, as applied to automatic and interactive processing of the aftershock sequence of the 2012 Sumatera (Ms=8.2) earthquake relative to the Reviewed Event Bulletin (REB) issued by the International Data Centre (IDC). The REB includes ~1200 aftershocks between April 11 and May 25 with (IDC) body wave magnitudes from 3.05 to 6.19. These aftershocks cover a slightly unusual V-shaped area. The cross correlation technique allows a flexible approach to signal detection, phase association and event building. To automatically recover the sequence, we selected sixteen aftershocks with $m_b$(IDC) between 4.5 and 5.0 from the IDC Standard Event List (SEL3) available on April 13. These events evenly but sparsely cover the area of the most intensive aftershock activity as recorded during the first two days after the main shock. After a superficial manual review these aftershocks were designated as master events. By no means is the selected set of master events the best or even an optimal one. There was no time and any specific procedure to estimate an event performance as a master and we used intuition based on previous experience. When the sequence is fully analyzed one may design a strict procedure as based on the optimal choice of important parameters. As an example, the number of stations with template waveforms has to be defined in the very beginning. In our study, waveform templates from only seven IMS array stations with the largest SNR estimated for the signals from the main shock were used to calculate cross-correlation coefficients over the entire period of 44 days with a time step equal to the relevant digitization rate. This rate varies among the stations. Approximately 1,000,000 detections obtained using cross-correlation were then used to build events according to the IDC definition, i.e. at least three primary stations with accurate arrival times, azimuth and slowness estimates. After conflict resolution between events with similar arrivals built by more than one master all qualified event hypotheses populated the automatic cross-correlation Standard Event List (aXSEL). The total number of distinct hypotheses was 4924. When all events matched by similar arrivals from at least one aftershock in the REB were excluded there were 2763 new hypotheses in the aXSEL. Many events in the aXSEL were built by arrivals in P-coda, with intensive water reflections as the major source of strong arrivals in coda. Since these reflections-based events have arrivals close in time to those from valid REB events, many of them were excluded after the comparison of the REB and aXSEL. Therefore, the total number of new hypothesis not matched by the REB is not the mechanistic difference between the aXSEL and REB. To evaluate the quality of new events in the aXSEL, we randomly selected a small portion (~10%) from 2763 events and analysts reviewed them according to standard IDC rules and guidelines. After the interactive review (a portion of) the final product of cross correlation was obtained – the interactive XSEL (XSEL). In the ideal case, the aXSEL should be identical to the XSEL, i.e. the automatic part of cross correlation processing should be tuned to provide the most comprehensive seismic catalogue without bogus events. Therefore, we have investigated the influence of many defining parameters (cross correlation coefficient threshold and SNR, *F*-statistics and *f-k* analysis, azimuth and slowness estimates, relative magnitude, etc.) on the aXSEL. We have constructed relevant frequency and probability density distributions for all detections, all associated detection, and for those which were associated with the aXSEL and final XSEL events. These distributions are also station and master dependent. When applied to the aXSEL from the beginning, these defining parameters would reject a vast majority of bogus events in the aXSEL not affecting the creation of valid event hypotheses.

Key words: array seismology, waveform cross correlation, REB, IDC, CTBT

## Introduction
In spite of its simplicity and long history of successful use in various seismological applications (e.g. Antsley, 1966; Geller and Mueller, 1980; Israelson, 1990; VanDecar and Crosson, 1990; Deichmann and Garcia-Fernandez, 1992) waveform cross correlation is a relatively new technique for massive computations of digital data from large archives with long time spans (Schaff and Richards, 2011). With a global distribution of mid- and large aperture array stations, many of them being the legacy of the arms race and extensive nuclear bomb testing era, the cross correlation technique gained an additional power: correlation with high quality waveform templates is superior to beam forming and thus provides a more effective detection tool (e.g. Gibbons and Ringdal, 2006, 2012; Harris, 2006, 2008; Harris and Pike, 2006; Ringdal *et al*., 2009; Bobrov *et al*., 2012).

The International Monitoring System (IMS) is likely the best example of a global array network. The International Data Centre is continually processing the IMS data, including arrays in North America, Asia, Europe, and Australia. Unfortunately, South America lacks any array station and the detection threshold is not distributed homogeneously over continents. The IDC has been producing automatic and interactive bulletins since 2000, with the Prototype IDC producing similar bulletin several years before the IDC was launched. Currently, there are around 400,000 events with seismic phases built by the IDC. The volume of waveform data available from the IMS arrays stations is determined by the time span of their operation, the number of channels, and sampling rate. Many array stations were built ten and more years ago. Some of them were reconstructed with the change in sensor configuration. A few stations are newly built. The number of sensors varies (with not all array stations built or even designed) from 6 (BRTR) to 42 (NOA). With a few exceptions, the continuous data at array stations are digitized at a rate of 40 samples per second during the past ten years.

The events in the Reviewed Event Bulletin cover the whole earth and accurately reproduce in the distribution of seismicity obtained by other global networks since the inception of seismological studies. All digital waveforms are archived together with the event information and the IDC historical set is ready for massive computations of cross correlation coefficient (*CC*). Taking into account the number of REB events and the overall length of digital waveforms, one needs a supercomputer power to estimate continuous *CC* time series for all array stations during the period of their operation.

For the purposes of seismology and geophysics, one of the most important tasks is to find all similar waveforms and thus co-located events. This set of repeating events from the same locations provides crucial information for tectonics, earthquake source physics (e.g. Yao *et al*., 2012), 3-D imaging of velocity structure and its evolution in time (e.g. Schaff and Beroza, 2004), and many others. By definition, the waveforms from repeating events recorded at the same stations have to be characterized by high cross correlation coefficients. This introduces an excellent opportunity to cluster historical events and implies a breakthrough in relative location procedure increasing the precision by one to two orders of magnitude (e.g. Schaff *et al*. 2004; Waldhauser and Schaff, 2008; Schaff and Richards, 2004, 2011; Selby, 2010).

For the IDC, there are additional tasks associated with cross correlation, which can provide a significant reduction in the global detection threshold and improvements in the statistics of event reliability. This invaluable gain for the Comprehensive Nuclear-Test-Ban Treaty is accompanied by a substantial reduction (by a factor of two and more) in the analyst workload as associated with the mandatory interactive review of automatic bulletins. The decrease is the detection thresholds is equivalent to the fall in magnitude threshold of nuclear test monitoring. The higher reliability of events built by cross correlation is guaranteed by a more precise characterization of valid arrivals and effective rejections of inappropriate arrivals (Bobrov *et al*., 2012). The analysts' workload depends on the quality of events and arrivals in automatic bulletins. As a result, the analysts need less time to check a smaller number of event hypotheses and detections of higher quality.

When based on waveform cross correlation, the seismic component of the CTBT monitoring system has to include a global grid of master events with high quality waveform templates at array stations. In seismically active areas, such master events are relatively easy to find. In aseismic regions, additional efforts are needed for construction and testing of synthetic master events based on optimal combination of empirical waveforms and/or synthetic seismograms. The global network of array stations has also to be completed. Before the CC monitoring system becomes operational one has to resolve a large number technical and scientific problems related to the optimal choice of the parameters best characterizing detections and events.

The IDC is obliged to find all seismic events, which one can retrieve from IMS data. Our experience with waveform cross correlation as an automatic detection and event building technique shows that, currently, the REB likely misses between 50% and 70% of events, some of them having valid arrivals at five and more IMS stations (Bobrov and Kitov, 2011; Bobrov, Kitov, and Zerbo, 2012; Bobrov *et al*., 2012). In addition, a relatively large portion of REB events does not demonstrate any cross correlation with neighbouring events, as expected from their relative locations and other parameters, and thus, such events are suspicious. These documented deficiencies of the REB have to be studied quantitatively at regional and global levels and a technical solution should be developed for a new automatic and interactive pipeline.

In this study, our principal objective is to assess the performance of waveform cross correlation as applied to automatic and interactive processing of the aftershock sequence of an extremely large earthquake measured by the IDC. Since this technique allows a flexible approach to all defining parameters, we investigate the effects of cross correlation thresholds, signal-to-noise ratio (SNR), relative magnitude and other parameters controlling the number of detections and the final list of events. In an attempt to optimize the analyst workload we test a reduced set of seven from nineteen IMS array stations reporting P- and Pn-waves from the studied earthquake. These seven stations are characterized by the highest SNRs measured from the main shock. The minimum number of associated stations for a valid event hypothesis is three, as required by the IDC event definition criteria (EDC).

We address the problem of optimal defining parameters for detection and event building as based on waveform cross correlation and propose simple recipes for conflict resolution between event hypotheses created by neighbouring master events. To demonstrate the power of cross correlation we selected the most complicated case for automatic and interactive processing at the IDC– the aftershock sequence of one of the biggest events ($M_s$(IDC)=8.2) in 2012, which occurred on April 11 near Sumatera. The REB for this sequence includes ~1200 aftershocks with ~350 during the first 16 hours. We aimed at repeating the REB using cross correlation and building as many new REB-ready events as possible, i.e. the events complying with the EDC. The REB was wholly repeated with an exception a few suspicious events, which are all small ones, and approximately 2760 new hypotheses were created. Due to the extraordinary large number of new hypothesis and limited human resources (analysts at the IDC are all busy with routine work) we had to carry out an "exit poll" exercise – just a small but randomly chosen fraction of all hypotheses was interactively reviewed, with the set of hypotheses from one master reviewed completely. At the same time, we have reviewed all events with six and seven (from seven used for cross correlation) defining stations because such events should never be missed by the IDC. Unfortunately for the REB and the current IDC pipeline, we have found five (!) seven-station events during the first 24 hours after the main shock, with the largest $m_b$(IDC)=4.95. This situation is inacceptable in nuclear test monitoring even for aftershock sequences of extremely large earthquakes. One can easily design an evasion scenario using this lacuna in monitoring.

In the current version of IDC processing, to be migrated in the REB an event has to match several strict quality (conditional probability) criteria. The most effective constraint is the number of primary stations reporting arrivals of primary seismic phases (i.e. Pg, Pn, P, PKP, PKPab, PKPbc). The minimum number of primary stations is three and the relevant detections have to be characterized by arrival time, azimuth and slowness residuals within predefined (station-dependent) uncertainty bounds. Using the underlying historical distributions of these three defining parameters (residuals) one is able to formulate a hypothesis on the probability for a given event to exist. This approach uses global distributions and thus is subject to global fluctuations in travel times and vector slowness. Waveform cross correlation is based on local distributions of all defining residuals, which have much smaller uncertainties. The sharper distributions allow formulating more reliable hypotheses on seismic events (Bobrov, Kitov, and Zerbo 2012; Bobrov *et al*., 2012), and thus build more valid events with a lower false alarm rate.

Theoretically, cross correlation detectors take the advantage of waveforms similarity what makes it superior to standard (STA/LTA) seismic signal detectors (Gibbons and Ringdal, 2006; Schaff and Waldhauser, 2010). In practice, the cross-correlation technique has proven to be a powerful tool for detection of similar signals (Harris and Pike, 2006; Gibbons and Ringdal, 2006, 2012). For array stations with many individual sensors at distances from a few hundred metres to tens kilometres, signals from close events have similar vector slownesses and should be synchronized at all individual channels, when synchronized at the reference channel. Signals from remote events are shifted by varying times relative to the reference channel (desynchronized) and thus are subject to destructive interference. This effect is best observed when a template waveform is correlated with the ambient seismic noise: the average cross correlation coefficient is of a few hundredths with the standard deviation of the same magnitude. Therefore, even a marginal level of cross correlation between two consistent signals around 0.2 usually provides an SNR high enough to detect any signal similar to that in the template waveform. This level of correlation between two time series is slightly counterintuitive for conventional physics, where the level below 0.7 is often considered as an insignificant one.

Using the waveform cross correlation technique, we have already recovered aftershock sequences of a mid-size Chinese earthquake (Bobrov, Kitov, and Zerbo, 2012) and a small event in the North Atlantic (Bobrov *et al.* 2012). Overall, we have found practically all REB events for these two sequences and a large number of new events meeting the REB criteria. For the event in China ($m_b$(IDC)=5.4), we found 36 new aftershocks by interactive review of only 45 hypotheses from the full set of 115. The limited human resources did not allow testing all hypotheses in the Chinese sequence and we carried out a comprehensive recovery experiment with a small event in North Atlantic ($m_b$(IDC)=4.2), where all REB and newly found events were iteratively used as master events before no new REB-ready events were found. In total, there were 26 new events build by cross correlation and reviewed by experienced analysts in addition to 38 REB events. In other words, we have obtained 67 per cent of new events. The quasi-ergodic properties of seismicity make it possible to extend the result of the comprehensive recovery to the global level. Therefore, one can expect the number of REB events to increase by a factor of 1.5 to 2.0 when the cross correlation technique is used instead of the current IDC processing pipeline.

Despite of very specific use of waveform cross correlation for the purposes of the CTBT, it also represents an important component in the formulation and testing of a new paradigm in seismology – the comprehensive use of historical (digital) data for precise location and characterization of global seismicity. Several studies within continents (e.g. Schaff and Waldhauser, 2005; Waldhauser and Schaff, 2008) have shown that the gain in location accuracy is from 10 to 100 times, i.e. the relative location can be reduced to a few hundred metres instead of 10 km. This is especially important in oceanic areas where dense local and regional networks are not available and most active tectonic movements occur. In this study, we demonstrate that a set of techniques based on waveform cross correlation significantly improve the completeness of the IDC catalogue in the zone of active seismicity and also improve location of all events (already and newly found) by associating them with a set of well-located master events.

## 1. Cross correlation, data, and master events

As in many cross correaltion studies (e.g. Gibbons and Ringdal, 2006; Schaff and Richards, 2011), we use a normalized cross correlation function despite it involves a (computationally) time consuming square root operation. As an alternative, one can use the representation based on squares of time series in the denominator as described by Gibbons and Ringdal (2012). Both time series must have the same sample rate. The notation $\omega_{N,\Delta t}(t_0)$ is used to denote the discrete vector of *N* consecutive samples of a continuous time function $\omega(t)$, where $t_0$ is the time of the first sample and $\Delta t$ is the spacing between samples:

$$\omega_{N,\Delta t}(t_0) = [\omega(t_0), \omega(t_0+\Delta t),\ldots, \omega(t_0+(N\text{-}1)\Delta t)]^T$$

The inner product of $\upsilon_{N,\Delta t}(t_\upsilon)$ and $\omega_{N,\Delta t}(t_\omega)$ is defined by

$$\langle \upsilon(t_\upsilon), \omega(t_\omega) \rangle_{N,\Delta t} = \sum_{i=0}^{N-1} \upsilon(t_\upsilon + i\Delta t)\, \omega(t_\omega + i\Delta t)$$

and the normalized cross-correlation coefficient, *CC*, by

$$CC[\upsilon(t_\upsilon), \omega(t_\omega)] = \frac{\langle \upsilon(t_\upsilon), \omega(t_\omega) \rangle_{N,\Delta t}}{\sqrt{\langle \upsilon(t_\upsilon), \upsilon(t_\upsilon) \rangle_{N,\Delta t} \langle \omega(t_\omega), \omega(t_\omega) \rangle_{N,\Delta t}}}$$

For a given waveform template at a given station, $t_0$ is fixed and the cross correlation coefficient at the elapsed time *t*, *CC(t)*, is defined as

$$CC(t) = \frac{\langle \upsilon(t + t_0), \omega(t_0) \rangle_{N,\Delta t}}{\sqrt{\langle \upsilon(t + t_0), \upsilon(t + t_0) \rangle_{N,\Delta t} \langle \omega(t_0), \omega(t_0) \rangle_{N,\Delta t}}}$$

where the term $\langle \omega(t_0), \omega(t_0) \rangle_{N,\Delta t}$ is the same for the whole *CC(t)* time series.

For a multichannel waveform associated with arrays stations, we consider two general ways to calculate $CC_t$. One can concatenate the template segments at all channels in a single time series with individual waveforms shifted by theoretical time residuals relative to the reference channel. It is also possible to calculate *CC* at individual channels and then average over all (qualified) channels with the relevant time shifts. For IMS arrays, both methods have their advantages and disadvantages, and we use the averaged *CC* in this study (Bobrov, Kitov, and Zerbo, 2012).

Cross correlation coefficient depends on the relative length of signal and correlation time window. When a signal is short and the window is wide, *CC* is likely underestimated because of dominating noise input. In a very narrow window, all signals look similar and *CC* is biased up. Because the Nuclear-Test Ban Treaty is a comprehensive one, the IDC has to focus at signals from near-surface events in the (body) magnitude range between 2.5 and 4.5. Seismic monitoring of the larger seismic events is globally comprehensive as the IDC catalogue demonstrates (Bobrov *et al.*, 2011). Smaller events are hardly seen even by regional networks. Therefore, all waveform templates in our study include several seconds of (P- or Pn-wave) signal and a short time interval before the signal (lead), which provides an additional flexibility in onset time.

For teleseismic and far-regional signals from low-magnitude events, one may expect the peak signal energy between 0.8 Hz and 6 Hz and the accuracy of *CC* estimates depends on relative frequency content of signals and background noise. A standard way to improve detection is to filter in the frequency band where the relevant SNR is the largest. The length of a given window also depends on its frequency band. For the low-frequency (BP, order 3) filter between 0.8 Hz and 2.0 Hz, the length is 6.5 s which includes 1 s before the arrival time. For the high-frequency filter between 3 Hz and 6 Hz, the total length is only 4.5 s. For Pn-waves, the window length is 11 s and does not depend on frequency. The Pn templates also include 1 s of preceding noise.

Figure 1 depicts a waveform template at station ZALV (blue line) as measured at a distance of 5992 km from one of the master events (0.222N 92.153E) we have used in the analysis of the April 11, 2012 Sumatera event (see Table 1). There are eight individual time series of ground motion measured at vertical channels with a sample rate of 40 samples per second. All individual waveforms are shifted by the corresponding theoretical time delays relative to the central (reference) channel as defined by the station/master event distance and back azimuth. These waveforms illustrate why cross correlation is superior to beam forming. Waveform cross correlation takes the full advantage of empirical time shifts between individual channels instead of the theoretical ones used in beam forming. At first glance, a few counts difference between theoretical and actual time arrival at a given channel is negligible. In reality, this difference introduces non-zero phase shifts, which cumulate in a substantial beam loss. Since we are looking for seismic events in the vicinity of master ones these empirical time shifts relative to theoretical arrival times are retained in all valid signals and absent in arrivals from remote events. Due to the high signal coherency, cross correlation is most effective for neighbouring events. This efficiency is most important for small events, which have weak signals mixed with microseismic noise.

All records in Figure 1 are filtered by a BP (order 3) filter between 0.8 Hz and 2.0 Hz, i.e. the window length is 6.5 s. For the sake of simplicity, we display only the segments used for cross correlation. There are 261 readings in each time series, and the arrival of P-wave in the master waveform at the reference sensor (the uppermost trace) corresponds to point 41. This template is moving along the continuous record at ZALV with a one-count time step, with the same theoretical time shifts between the channels in the record. When the master template is cross correlated with microseismic noise, the estimated cross correlation coefficient is very low. At some point, the template reaches a similar signal and *CC* increases above the noise level. There is a time point where *CC* has a local peak, say, within a ±4 s time interval around the peak. An example of a "slave" waveform having a local *CC* peak with the master event is shown by red line in Figure 1. This slave event corresponds to another master event used in our study which was approximately 55 km away from the first master according to its IDC location (0.5845N 92.495E). Both events had similar $m_b$(IDC) magnitudes of 4.89 and 4.63, respectively, and their absolute amplitudes at ZALV are almost identical. The arrival time of the slave signal at the reference sensor of ZALV in Figure 1 corresponds to the highest *CC*=0.632. (Since we introduced a 1 s pre-signal interval in the master waveform, the found slave signal also lags by ~1 s behind the first point of the slave record.) One may observe the changing difference between waveform shapes and time shifts on individual channels. Specifically, both wave trains are poorly synchronized beyond the first two seconds of the signal. This is due to the path difference, which is likely more prominent for reflected/refracted waves.

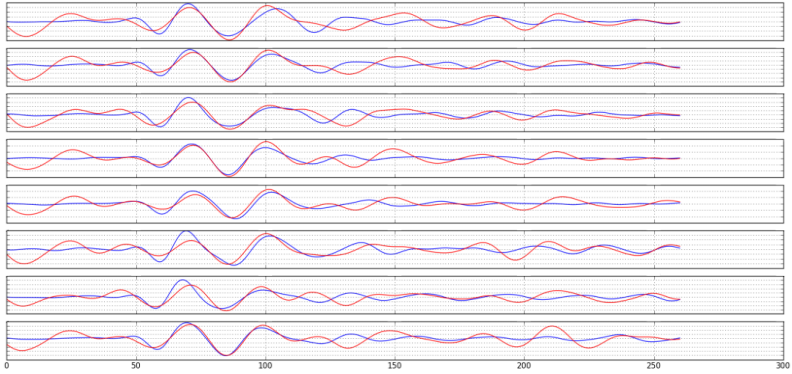

**Figure 1.** Cross correlation of master (blue) and slave (red) waveforms at station ZALV. *CC* is calculated for the first point of the slave signal. *CC*s are first estimated for each of eight channels and then averaged. The frequency band is 0.8 Hz to 2.0 Hz and the corresponding time window length is 6.5 s (261 counts). One may observe the changing shifts in arrival times between the master and the slave.

In this study, we first calculate *CC* for each individual channel and then average these *CC*s over the channels to obtain an aggregate *CC* estimate for a given time point. Since we correlate and then average two multichannel ground motion waveforms, i.e. the template and the whole record, the set of several channels is replaced by one *CC*-trace. This trace is used to detect signals with a standard STA/LTA detector and to estimate the peak *CC* for these signals. The length of short-term and long-term windows is flexible; we use 0.8 s and 20 s, respectively. (One may also apply different detectors to the aggregate *CC*-trace and to the multichannel *CC*-traces, e.g. *F*-statistics.) Without loss of generality, we ignore all signals which have |*CC*|<0.2 or STA/LTA<2.5. According to our previous studies, these are conservative thresholds rejecting a vast majority of inappropriate signals but still allowing some bogus signals in noisy environment. We need these bogus signals in order to determine statistically justified decision thresholds for all master/station pairs. These thresholds may balance the rate of missed valid signals and that of bogus signals.

One can also estimate *CC* with all channels of the template concatenated in one record and cross correlated with the slave record aligned in the same way. In this case, the empirical time shifts of individual channels are also retained and cross correlation is more effective then beam forming. The alignment gives *CC* estimates similar and even slightly superior to those obtained by averaging of individual traces but suffers from data quality – spikes, steps, missing data and channels introduce spurious signals in the *CC*-trace (Bobrov, Kitov, and Zerbo, 2012). This is the reason we use the averaging technique with tapering of problem channels as a more reliable *CC* estimation procedure.

At this stage, all individual *CC*-traces are retained for further analysis. They have the same sampling rate and time shifts relative to the reference sensor as the underlying waveforms. (In that sense, all signals obtained by cross correlation with a given master are well pre-filtered in azimuth and scalar slowness.) The important difference is that the *CC*-traces are normalized and thus do not depend on the relative master/slave size. In addition, the input of noisy channels is heavily suppressed on the *CC*-traces since microseismic noise does not correlate with master waveforms. According to Gibbons and Ringdal (2012), these advantages allow obtaining more reliable estimates of pseudo-azimuth and pseudo-slowness (i.e. those not measured in degrees and s/degree) by standard *f-k* analysis even for very weak signals measured at a few channels. In order to characterize the quality of signals detected with cross correlation coefficient we also estimate *F*-statistics for the multichannel *CC* records. In this study, we compare results obtained using regular waveforms with those from the corresponding *CC*-traces.

When a signal is detected by cross correlation, we calculate all standard IDC attributes such as amplitude, period, azimuth, scalar slowness, SNR, quality class, etc. For a given source/station distance, one can calculate body wave magnitude using amplitude and period. Magnitude is a helpful dynamic parameter for global association of seismic phases – only signals in a predefined magnitude range can be associated with one source. Cross correlation provides another statistically powerful dynamic parameter for phase association. To characterized relative sizes of two events having a cross correlation coefficient above some threshold at a given station, we propose to use the ratio of their $L^2$-norms, $|x|/|y|$, where $x$ and $y$ are the vectors of data for the slave and master event, respectively. In other words, we use the ratio of RMS amplitudes calculated in the same (correlation) window. The logarithm of the ratio, $RM = \log(|x|/|y|) = \log|x| - \log|y|$, is essentially the magnitude difference or the relative magnitude of two events. This difference has a clear physical meaning for close events with similar waveforms, i.e. for events with a higher (e.g. >0.2) cross correlation. It does not always work well for events at large ranges because of the difference in propagation paths and likely in source functions. In this case, standard magnitude scales work better. We have already shown that the RM is a better dynamic parameter for discrimination between genuine and dynamically inappropriate arrivals for a given event at several stations (Bobrov, Kitov, and Zerbo, 2012).

Having a detector based on cross correlation and an extended set of signal attributes, including the relative magnitude, one can process routine IMS data and associate signals with events, i.e. to build event hypotheses. The association process is much simpler than that realized in the Global Association algorithm (Coyne *et al*., 2012) currently used at the IDC. Since we are looking for seismic sources close (0 km to 70 km) to a known set of master events a local association procedure, *LA*, is feasible (Bobrov, Kitov, and Zerbo, 2012). The principal assumption of the *LA* is that the travel times from a given master event to IMS stations are a good first approximation for any slave event, which (by definition) is close to the master. Therefore, actual travel time residuals for any slave event have to be within a few (say, 6) seconds from those predicted by the master. By projecting actual arrival times measured at several IMS stations back to source with the relevant masters/station travel times one can obtain a set of origin times. For a true REB-ready slave event, three or more origin times associated with different stations have to group within a short interval (e.g. 8 to 12 s) and all attributes of the relevant signals (azimuth, slowness, relative magnitude) should be within station-specific uncertainty bounds (Coyne *et al*., 2012).

We have already reported results for two relatively weak aftershock sequences, where one master event was enough to cover the whole aftershock area. Here we exercise an event with aftershocks covering more than 250,000 km$^2$. This requires quite a few master events spaced by 70 km to 100 km, i.e. each master has to cover a circular area 50 km to 70 km in radius in order to avoid any blind zones. Such masters have to be preselected, when historical seismicity is available for the studied area, or to be taken directly from the fresh aftershock sequence. For the IDC, the latter supposes an expedite review of some earlier events available in the automatically built Standard Event List (SEL3).

This study was initiated by a very big earthquake near Sumatera and thus was chiefly spontaneous in its initial phase. This explains why there was no specific procedure for master events selection. We had to design the procedure and select master events under time pressure in the absence of specific experience. As a result, the set of masters might not be an optimal one for this aftershock sequence. Intriguingly, all selected masters have demonstrated a good performance. Hence, when a truly optimal set of master events is used, one could expect more new aftershocks found by cross correlation. In a comprehensive mode, when all (qualified by the number of stations and signal SNRs) aftershocks are used as masters, the cross correlation technique may double the REB content.

The main shock of the Sumatera April 11, 2012 earthquake was measured by the IDC, which located the event 2.294N and 93.035E with the origin time 08:38:32.23 (UTC). The solution was fixed to surface. Body wave magnitude $m_b$(IDC)=5.82 is small judging by $M_s$(IDC)=8.2. This effect of $m_b$ underestimation for larger events (also relative to other global seismological centres, e.g. the ISC) is well known and is associated with the length of time window for peak amplitude and period estimation. There were 162 associated and 60 defining phases including two Pn-arrivals and fifty two P-arrivals. Figure 2 shows the location of the main shock (green square) and the distribution of aftershocks between April 11 and May 25, 2012. We have processed waveform data at eighteen IMS array stations during these 44 days and compared the relevant REB and automatic XSEL.

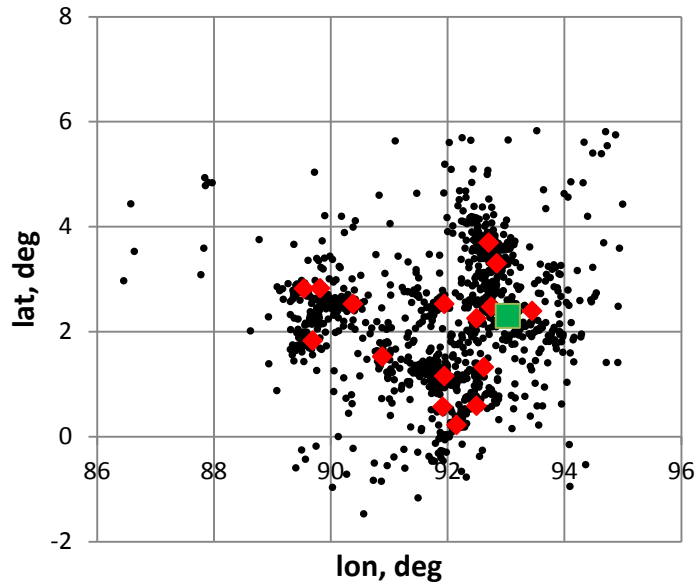

**Figure 2. Distribution of the REB events (black circles) related to the Sumatera earthquake (green square) between April 11 and May 25, 2012. Sixteen master events are shown by red diamonds. They were preselected using the SEL3 available on April 13 and reviewed by an experienced analyst. The events at the far periphery of the aftershock zone are likely mislocated due to low location accuracy of 3- to 5-station events and probable mis-timing/mis-association of low-amplitude arrivals.**

Figure 2 demonstrates that one may consider the area of the aftershock zone as constrained by a square between 88.0N and 94.0N, and 1.0S and 4.0N. Thus, the principal area is approximately 500x500 $km^2$. There are many events sparsely distributed far away from the main aftershock zone, which has a slightly unusual V-shape. The events to the west, north and south of the zone are likely mislocated by hundreds kilometres. Some of the events to the east of the zone are likely mislocated but may also be related to the seismically active zone beyond 95E. Therefore, we have limited our study to the REB events west of 95E. The relevant REB includes 1181 aftershocks which we would like to find using waveform cross correlation. We also expected to find from 600 to 1000 more REB-ready events, i.e. the XSEL has to be 50% to 70% larger than the REB. In any case, the first task was to define the set of quality master events. Considering the principal aftershock area described above and our previous finding that the spacing between a master and the events to be found has to be less than 50 to 70 km we decided to limit the number of masters to 16. If this assumption would have been wrong, we could easily extend the list of masters.

There were no REB events in the area of the studied earthquake and its aftershocks, and no masters could be preselected. Therefore, we had to select a high quality and comprehensive set of master events from the aftershocks during the first day after the main shock. By IDC rules, SEL3 for a given day has to be ready six hours after the data day. There was no plan to recover the aftershock sequence before April 11. The decision to start cross correlation analysis in parallel to the routine automatic processing was made on April 12. On April 13, we sought though SEL3 for days 1 and 2 after the main shock. Without any formal procedure we selected sixteen (approximately) evenly distributed events with $m_b$(IDC) between 4.34 and 5.06 within 24 hours after the main shock. Table 1 lists: the REB internal indices uniquely defining the master events, ORID, coordinates, the number of phases defining the relevant REB solutions, ndef, and body wave magnitudes, $m_b$. (Other two parameters are discussed later in the text.) The number of defining phases (mainly P-waves) guarantees that the locations and magnitudes of all masters are accurately estimated. Figure 2 shows the distribution of master events as reported by the REB. This distribution is slightly different from that obtained by the expedite review on April 13. Since cross correlation is based on relative positions of masters and slaves, we ignore the difference between the REB and these preliminary locations of the master events.

**Table 1. Sixteen master events selected from SEL3 and reviewed by an experienced analyst.**

| ORID | Lat, deg | Lon, deg | ndef | $m_b$ | SNR_AV | # found events |
|---|---|---|---|---|---|---|
| 8597212 | 0.585 | 92.495 | 41 | 4.89 | 21.5 | 2074 |
| 8597277 | 3.301 | 92.841 | 47 | 4.52 | 55.3 | 2391 |
| 8598014 | 1.526 | 90.881 | 57 | 4.69 | 32.4 | 2124 |
| 8598080 | 2.529 | 91.945 | 52 | 4.34 | 17.3 | 2307 |
| 8598208 | 0.222 | 92.153 | 45 | 4.64 | 20.6 | 2133 |
| 8598640 | 1.321 | 92.621 | 67 | 4.81 | 45.4 | 2166 |
| 8599430 | 2.820 | 89.544 | 44 | 4.67 | 30.6 | 2149 |
| 8602152 | 2.469 | 92.750 | 67 | 5.06 | 186.7 | 2636 |
| 8602260 | 2.394 | 93.451 | 67 | 4.94 | 208.4 | 2487 |
| 8602336 | 2.254 | 92.498 | 47 | 4.73 | 14.6 | 1898 |
| 8602568 | 1.829 | 89.695 | 74 | 4.99 | 76.1 | 2309 |
| 8603230 | 2.828 | 89.819 | 37 | 4.61 | 10.8 | 1785 |
| 8604933 | 1.152 | 91.944 | 42 | 4.50 | 16.9 | 1748 |
| 8605332 | 2.528 | 90.380 | 70 | 4.96 | 29.6 | 2379 |
| 8605418 | 0.570 | 91.920 | 47 | 4.59 | 16.2 | 2303 |
| 8606054 | 3.693 | 92.705 | 52 | 4.46 | 37.9 | 2068 |

To cook a good set of master events one needs to select the best waveform templates. The master waveforms for cross correlation have to be clear and representative in sense of source function and focal mechanism. The first requirement is formally quantified by signal-to-noise ratio – clear signals have very high SNRs. The second requirement has to balance the whole range of empirical signal shapes and their dependence on magnitude. Seismic events with large magnitudes are characterized by an enhanced content of low frequencies (low corner frequency of source spectrum) and small events are usually manifested by short and higher frequency signals as associated with generally higher corner frequencies. From our previous experience, the choice of master (IDC $m_b$) magnitudes between 4.5 and 5.0 might resolve this issue. Such mid-size masters "see" low magnitude (say, two units of magnitude smaller than the relevant master) and much bigger events, i.e. they provide a good shape approximation for both extremes, which results in relatively large cross correlation coefficients. In almost no case, real master events can be smaller than 4.0 since their signals would not have high enough SNRs at five to ten stations.

In order to select a self-consistent set of waveform templates for the sixteen masters we have inspected P-wave SNR at all IMS array stations from the main shock. Table 2 lists SNRs for 20 IMS array stations in descending order. The highest SNR belongs to MKAR and the lowermost to PETK, which was excluded from processing because of low sensitivity to the Sumatera aftershocks. Station ILAR is also excluded from the list since the event-station distance is on the edge of P-wave existence and some aftershocks may be in the core shadow zone. Thus, from the set of 18 stations, we had to select a subset with high SNRs, which is common for all masters. It had also to balance the workload of automatic/interactive processing and the performance of cross correlation including reliable event location/building; as a mandatory requirement this small subset of stations had to provide a good azimuth distribution. Fortunately, the top seven stations in Table 2 fit these requirements and we initially processed only waveform templates for the following stations: MKAR, WRA, CMAR, SONM, ASAR, ZALV, and GERES. An analyst reviewed all candidate signals and corrected onset times where necessary. Finally, we obtained the set of 16 master events and each of the masters had 7 waveform templates at the best array stations for the main shock. We assumed that the relative amplitudes of signals (and SNRs) from aftershocks at 18 stations follow the same general trend as in Table 2.

In order to assess the overall quality of templates, we have averaged all SNRs for each master, as listed in Table 1 (*SNR_AV*). The highest average SNR is 208.4 and the lowermost is 14.6. This is a significant scattering and one might select a better set of masters with higher *SNR_AV* in some optimal procedure. We have learned this lesson.

Altogether, our cross correlation detector has found 998773 arrivals at 18 primary array stations during the studied period. (Notice, the input of ESDC is only 6 detections) The selected subset of the best seven stations has given 621111 detections (62.1%). Table 2 lists the distribution of detections over stations. The highest sensitivity is associated with MKAR, which detected 170827 P-wave arrivals. This is an expected result since MKAR has by far the highest SNR among all stations. Unexpectedly, station NOA gave only 2361 detections while FINES and ARCES, having the very same SNR associated with the main shock, reported 31524 and 22392 arrivals, respectively.

Figure 3 depicts the frequency distribution of *CC* for all detection obtained with the seven-station master events (autocorrelation is excluded). There are two distinct segments associated with low and high *CC*. Between 0.2 and 0.6, the distribution is a quasi-exponential one, both for negative and positive *CC*s, with the exponent of ±5.94 and a very high coefficient of determination ($R^2$=0.996). We have reported similar behaviour for the aftershock sequence of the Chinese earthquake (Bobrov, Kitov, and Zerbo, 2012). For an aftershock sequence evenly distributed in space and having a standard magnitude recurrence curve (a power law distribution), the exponential fall with *CC* may reflect a quasi-exponential *CC* dependence on master/slave distance. For very close events with high *CC*s, the difference in shape and the absence of smaller events with low-SNR signals may play the leading role in the faster roll-off of the distribution.

Obviously, low-*CC* detections have a smaller probability to be associated with events: they have a higher uncertainty in arrival time, azimuth and slowness, have a higher probability to be noise-related, and may just miss the third station to match the REB criteria. The portion of associated arrivals is expected to rise with *CC*. The fall in the absolute number and the increase in the portion may result in a peak *CC*, which is also station dependent. For the purposes of event building, one may define for each master/station pair the probability for detections with a given *CC* to be associated with an REB event. This is an effective procedure to control the rate of false event hypotheses.

**Table 2**. IMS array stations, main shock – station distances, arrival times, signal-to-noise ratios, and number of detections.

| STA | Δ, º | Arrival time, s | SNR | # detections |
|---|---|---|---|---|
| MKAR | 45.30 | 1334134011. | 248.0 | 170827 |
| WRA | 46.16 | 1334134018. | 169.2 | 104954 |
| CMAR | 17.08 | 1334133751. | 113.5 | 68212 |
| SONM | 46.81 | 1334134023. | 99.8 | 101104 |
| ASAR | 47.40 | 1334134028. | 87.6 | 60634 |
| ZALV | 51.93 | 1334134062. | 84.6 | 48742 |
| GERES | 81.26 | 1334134250. | 77.1 | 66638 |
| KSRS | 47.41 | 1334134029. | 73.6 | 27722 |
| GEYT | 47.76 | 1334134031. | 67.2 | 30430 |
| FINES | 77.13 | 1334134228. | 56.2 | 31524 |
| BRTR | 65.31 | 1334134157. | 54.8 | 66142 |
| ARCES | 80.12 | 1334134243. | 54.6 | 22392 |
| NOA | 84.03 | 1334134263. | 54.1 | 2361 |
| USRK | 54.09 | 1334134080. | 48.4 | 66447 |
| AKASG | 71.87 | 1334134195. | 45.9 | 19430 |
| TORD | 90.79 | 1334134298. | 43.2 | 31109 |
| MJAR | 53.80 | 1334134077. | 38.0 | 37077 |
| ESDC | 93.94 | 1334134314. | 33.5 | 6 |
| ILAR | 100.29 | 1334134341. | 18.6 | - |
| PETK | 73.17 | 1334134207. | 10.0 | - |

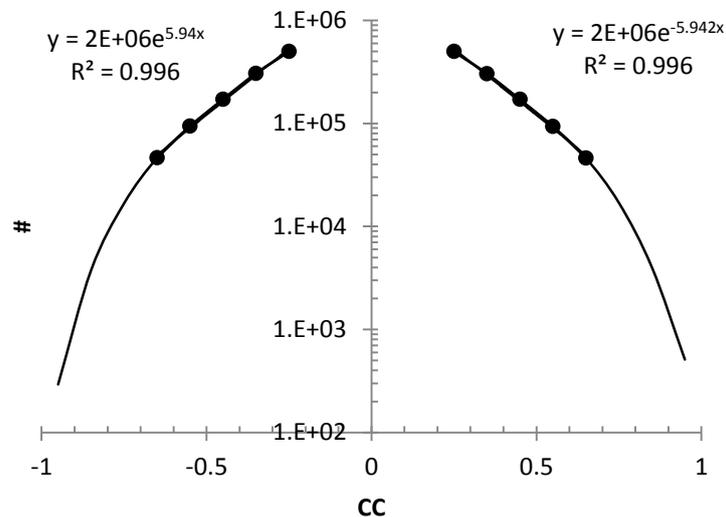

**Figure 3.** Frequency distribution of *CC*s as obtained from all 16 masters. In the ranges between 0.2 and 0.6 and from -0.2 to -0.6, the distributions are practically identical and both exponential. Beyond ±0.6, the fall is steeper than exponential. Autocorrelation is excluded.

Having absolute arrival times for all detections it is possible to calculate approximate origin times by subtracting corresponding master/station travel times. For each master event, one can find all groups of three or more stations with origin times within 8 s. These groups create zero-level event hypotheses. Apparently, theoretical (master/station) travel time depends on the accuracy of master location. At the same time, the slave event can be 50 and even more kilometres far from the master. To compensate the influence of the master mislocation and the master/slave distance on the origin time we have introduces a dense grid of 18 virtual masters: six at distances 20 km and twelve at 40 km from each master event. Effectively, we recalculate travel times for each grid point and subtract them from the same arrival times. Therefore, there are 19 different hypotheses on the origin times and we choose the one with the largest number of stations and the lowermost origin time RMS residual. In other words, we select the tightest group of origin times among the most station-populated groups. The average origin time is considered as the event origin time for the best hypothesis.

The above procedure is similar (but not equivalent) to grid search in the well-known double difference location. To improve the onset times for the grid points we could recalculate *CC* with appropriately changing theoretical time shifts between channels for a given station. This procedure preserves the empirical time shifts between channels, which are the most important advantage of cross correlation. These additional calculations would require extra computer resources but could definitely bring more valid detections with low *CC*. Under the CTBT framework, such detections are of the highest priority.

When the set of zero-level hypotheses based on origin times is built, we resolve a number of conflicts between the arrivals in the same hypothesis using amplitude, azimuth and slowness residuals. All survived hypotheses join the set of REB-consistent hypotheses for each master event. As mentioned above, pseudo-azimuths and pseudo-slownesses are estimated by *f-k* analysis on *CC*-traces and used to remove inappropriate arrivals. For the original waveforms, the deviation of the estimated parameters from their theoretical values has to be within station-dependent bounds. For the pseudo-azimuth and pseudo-slowness, we have to determine their own uncertainty bounds expressed in pseudo-degrees and pseudo-second-per-degree, respectively. The deviations are measured from theoretical azimuth and slowness estimated for a given masters: autocorrelation gives zero deviations from the master. In this study, we formally introduced the following uncertainty bounds expressed in pseudo units: valid azimuth residuals have to be within $\pm 20^o$ and valid slowness residuals are expected within $\pm 2$ s/deg for all arrays. Thus, the azimuth/slowness test is a simple one, which we actually partly implemented at the detection stage, i.e. all detections are already within some global uncertainty bounds relative to station/master vector slownesses. At the stage of local association, we are planning to apply station dependent limits and to make the association procedure more flexible. The estimation of these station limits from the XSEL is one of our tasks.

There should also be an overall consistency between dynamic parameters of the arrivals associated with one event. We have introduced the relative magnitude as a measure of source strength relative to master events. From our experience, we have estimated the global level of allowed deviation of a station relative magnitude, *dRM,* from the average value for a given event, i.e. station magnitude residual. For the Chinese aftershock sequence, ninety per cent of these residuals reside within $\pm 0.4$ units of magnitude. In this study, we have defined the magnitude residual threshold as 0.7: no absolute deviation from the network average relative magnitude is above 0.7. At the same time, we expect the distribution of *dRM* in the XSEL to have less than 5% of arrivals beyond $\pm 0.4$. The relative magnitude might be a more efficient parameter to suppress the creation of bogus events for regular seismicity than for aftershocks. Two or more arrivals close in time at a given station may have similar RM but belong to one (the first arrival and coda arrivals) or a few aftershocks.

When all azimuth-slowness-magnitude conflicts between arrivals in a given eight-second-wide interval are resolved there might be more than one event with practically the same origin time. This may often happen in the initial stage of a large aftershock sequence when the rate of aftershocks is very high. We have considered such a situation and allowed new hypotheses to be formulated when the number of not associated (e.g. by RM or azimuth) arrivals in the window is 3 or more. It might also happen that the arrivals out of the window may be associated with the leftovers within another eight-second window. We are trying to associate any arrival with a potential REB event considering its kinematic and dynamic attributes. No arrival for a given master can be associated with two or more events.

Table 1 lists the number of aftershocks found by 16 master events. It varies from 1738 to 2636, but concentrates around 2100. At this stage, we obtain a preliminary automatic event list, aXSEL, which has to be as close to the final (interactive) XSEL as possible. The final aXSEL is not a mechanistic sum of all individual masters, however. Only the closest master is able to find the smallest slaves, but bigger aftershocks can be built by many masters. There are two effects creating multiple solutions and thus working against the simplicity of local association. First, the correlation distance depends on the vector slowness difference between master and slave. Figure 1 shows that there is just a small shift in relative arrival times at individual sensors. Station ZALV is ~6000 km far from the master event and the apparent velocity of the P-wave along the surface is more than 25 km/s. Therefore, the largest difference between the arrival times at individual sensors is less than 1.0 s since the array aperture is less than 20 km. For the slave event, the scalar slowness differs by a small fraction and the azimuth is essentially the same. As a result, the time shifts of the slave signals at individual sensors of ZALV relative to those from the master event are marginal and cross correlation does not suffer much. This effect is most important for events at the same great circle with station. Since we use a set of stations with good azimuth coverage, the effect of the great circle is suppressed by network processing: origin times in different directions scatter beyond 8 s when master and slave are 100 and more km away. In any case, one should be very careful when using three stations in the same azimuth to build an event hypothesis.

Second effect resulting in multiple event hypotheses is associated with larger seismic events. Their waveforms are much longer than the correlation time windows and the waveform templates may find peak correlation coefficients not in the beginning of the records but later in the wavetrains. This may happen even for signals arrived from different azimuths. In a way, the correlation technique is blinded by bigger events like a night-vision system is blinded by day-light. A simple way to avoid the multiple hypotheses is to separate all bigger events by their relative magnitude and review them interactively. There is no case one can miss the bigger events. Therefore, this issue is not critical for the CTBT.

There are 137274 detections left after the first round of phase association and event building. This is 22 per cent of the total number of detections (621111) obtained by seven stations and sixteen master events. These not-associated phases are not necessarily bogus detections. A larger portion of them could build seismologically valid two- and one-station events, which are not considered in our study because of the EDC. Apparently, with event magnitude increasing from very low level, its signals would first be detected by the most sensitive station (MKAR likely plays this role for the Sumatera aftershocks) and then by two best stations. Only when detected by three IMS stations it can be qualified as an REB event and thus be listed in the aXSEL.

For the total number of 34957 built events (Table 1), one has 3.96 phases per event. Therefore, the average event built by cross correaltion is a 4-station event. Actual numbers are: 17202 3-station events, 8555 4-station, 5073 5-station, 2806 6-station, and 1321 7-station events. Bigger aftershocks have a higher probability to be built by many master events and their best solutions should be highly reliable. In many cases, three-station events are bogus ones because we intentionally put our thresholds rejecting the (kinematically and dynamically) inappropriate detections too low.

We have developed and tested a tentative procedure to resolve conflicts between events with similar arrivals, which are built by different masters, in order to leave only one best hypothesis in the aXSEL for the further interactive review. To begin with, we check the arrival times at all stations. Obviously, origin times for various masters may vary in a relatively wide range because of travel time differences, but for the same (in physical terms) event all masters have to find the same physical signals, which belong to the sough slave event. When several (from 2 to 16) master events have similar arrivals at the same station, i.e. close (±4 s) arrival times and magnitude estimates (master magnitude + *RM*), one can consider them as generated by the very same source. Between all hypotheses involving similar arrivals we select those with the highest number of defining stations. This is a natural choice since all master events have the same stations. (To populate the master event list, one also should guarantee that signal's SNR at a given station should be similar for all masters. We did not apply this criterion when selected the master events and this might affect the aXSEL.) If two or more master events have the same number of defining stations, we give the priority to that with the highest cumulative (absolute) *CC*. This completely resolves any conflict between masters having the same set of arrivals and only one event hypothesis left in the aXSEL. All arrivals associated with other masters for the same event are removed from the list of detections. Because of the grid search, we also estimate the slave event location relative to its master event.

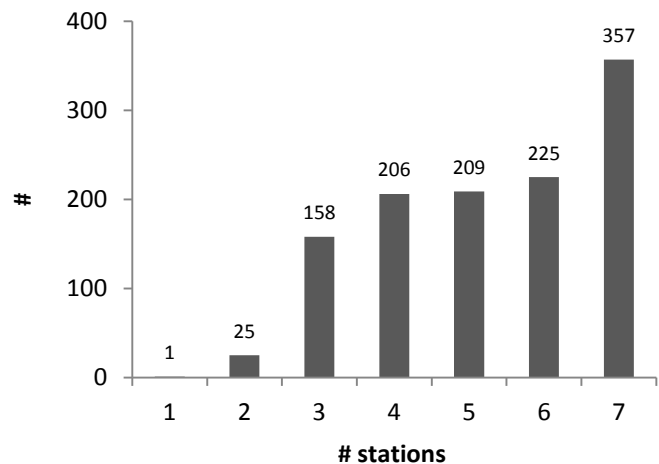

**Figure 4. Frequency distribution of the number of aftershocks in the REB with a given number of stations (from 7 involved).**

Before we discuss various features of the aXSEL for the period between April 11 and May 25, we would like to present some important properties of the relevant REB, which includes 1181 aftershocks and one main shock. Figure 4 displays the frequency distribution of the number aftershocks with a given number of stations from those seven used in the master events. There are only 26 events having less than three stations from the seven. (No REB events were re-reviewed in this study but will be analyzed in due course.) Moreover, 997 (84%) events have four and more stations, i.e. are well constrained for the REB. This means that our assumption on the highest sensitivity of the stations chosen for cross correlation is practically precise. Since one needs only three primary stations for an REB event the value added by other 11 stations from Table 2 is almost negligible. Extra stations might add some location and magnitude accuracy, which are very important for the REB. However, cross correlation may guarantee very accurate location, small confidence ellipse and a precise magnitude estimate because all slave events are close to their masters.

Moreover, smaller events are much better located as they have to be very close to their masters. This finding has to be used in designing the procedure for master events selection – a few stations with highest sensitivity can make the job. We skip here the discussion on the input of regional 3-C stations, which are always helpful for the smallest events.

Figure 5 depicts an REB recurrence curve for the studied aftershock sequence, as expressed by the number of events in 0.2-wide bins. There is a clear interval between magnitude 4.1 and 5.1 with a power law distriution. It is possible to consider $m_b(IDC)=4.2$ as the catalog completeness threshold, i.e. the REB progressively misses more and more events with decreasing magnitude below 4.2. The principal purpose of the XSEL is to move this threshold as low as possible. At least, it would be helpful to completely fill the bin between 3.8 and 4.2 (black diamonds represent the final automatic XSEL as discussed later on), but we have also found many new events with magnitude above 4.2.

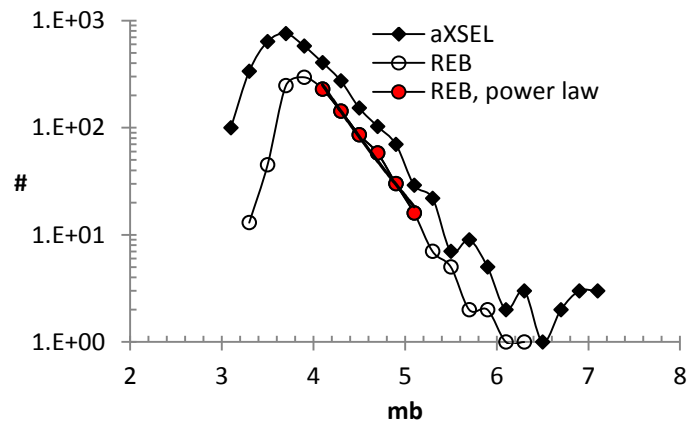

**Figure 5. The recurrence curve for aftershocks as obtained from the REB and automatic XSEL. For the REB, a power low distribution is observed between 4.1 and 5.1. For the aXSEL, the completeness threshold is 0.3 to 0.4 units of magnitude lower.**

The entire phase association/event building procedure was developed during the first days after the main shock and we had to test it using fresh data. Before analysts started the interactive review of the SEL3 for Aprill 11, we had created a preliminary aXSEL for the first three days. Among hundreds hypotheses, this three-day aXSEL also included 25 new events with seven defining phases (stations) absent from the SEL3. This finding was of importance for monitoring and all 25 hypotheses were included in the interactive review by force. (The IDC should not miss big events even and specifically after the major earthquakes.) Twenty event hypotheses were found associated with reflected/refracted P-waves (P-wave coda) from very big aftershocks. In the first version of the *LA*, we allowed colocated events spaced by 12 s in origin time. In terms of physics these are imaginary events with properties practically equivalent to those in the relevant true events. In that sense it is very difficult to distinguish between the imaginary and true events except the former should be observed later in time. At the same time, an evasion scenario is possible when an explosion is mixed with P-wave coda of a large earthquake. Hence, the IDC should not suppress the events based on P-coda. In the initial stage of our investigation, we decided to remove all colocated events within 40 s in order to reduce the workload for the post-mortem interactive analysis. But we remember that no events should be neglected in routine cross correlation processing.

There were five true events absent in the SEL3 and thus potentially missed by the REB. Table 3 lists their origin time, location, number of defining and associated phases, and magnitudes. Not surprisingly, analysts confirmed all phases found by the automatic cross correaltion detector. Figure 6 shows two events with six defining arrivals on array stations. These events had the largest and smallest magnitude in Table 3. The event in the upper panel occurred in ~13 minutes after the main shock. The hours after the main shock are charaterized by very high flux of arrivals from the same region which is hardly resolved by the *GA*. There were many six- and five-station events for the first day which were not reviewed in the beginning. Some of them were analyzed later.

There is a question whether the number of stations in master events is adequate to the task of an effective recovery of the aftershock sequence? To address this question, we have compared the total (summed over all masters) number of event hypothesis with 7 and 18 stations: 34957 and 43489, respectively. (Only signals with SNR>5 are used in the relevant waveform templates.) Figure 7 shows the gain obtained by each master event, which is 24% on average. Smaller (in magnitude) masters gain less because they miss many templates with low SNR. The actual value added by extra 11 stations may be much less than 24%, however. Figure 8 displays the number of hypotheses as a function of the number of stations. We see that the number of 4-, 5-, and 6-station event hypotheses is almost the same for 7- and 18-station sets. For example, the number of 5-station events is 5073 and 5117, respectively. This assumes that practicall no station from the 11 can add a phase to a 5-station event built only from the 7 stations. In other words, when the sixth and seventh best stations are absent the probability of the eighth and further station to be associated is negligible. It is easier to find the $6^{th}$ or the $7^{th}$ station, however. The number of 3-station events increses substantially, but the rate of bogus events might increase proportionally. In interactive analysis, the third station often has correct attributes from the correlation point of view but analysts can not find a standard signal and attributes obtained by standard (waveform based) *f-k* analysis.

**Table 3. New events found for the first day after the main shock as based on 7-station hypotheses.**

| Date | Origin time | Lat, deg | Lon, deg | Depth, km | nass | ndef | mb |
|---|---|---|---|---|---|---|---|
| **04/11/2012** | 8:51:33 | 2.44 | 92.59 | 0 | 7 | 7 | 4.95 |
| **04/11/2012** | 9:32:02 | 2.20 | 93.09 | 0 | 8 | 7 | 4.29 |
| **04/11/2012** | 10:05:05 | -0.26 | 92.59 | 0 | 7 | 7 | 4.06 |
| **04/11/2012** | 13:42:53 | 2.10 | 93.62 | 0 | 17 | 17 | 4.56 |
| **04/11/2012** | 16:25:48 | 3.31 | 92.71 | 0 | 7 | 7 | 3.88 |

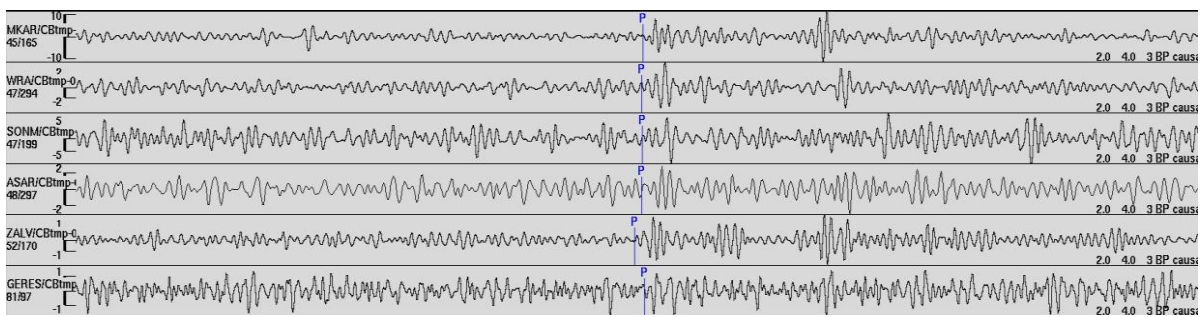

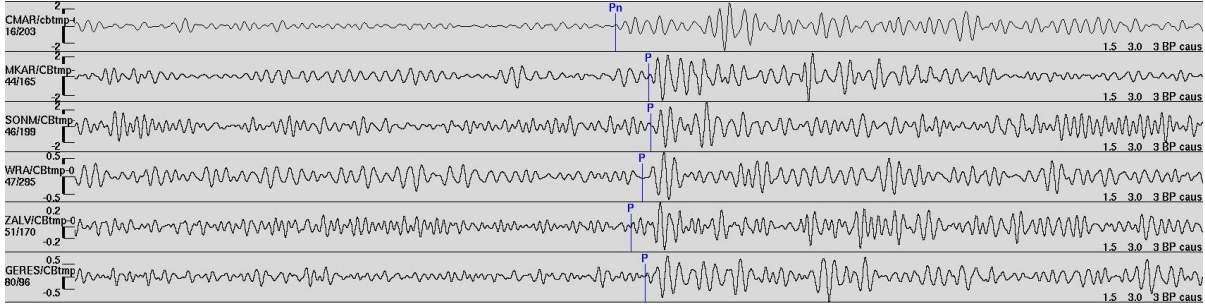

**Figure 6. Two new events with seven defining stations in the aXSEL. Original waveforms are filtered (filters are given on the right side) and centred on arrivals. Only array stations are shown.**

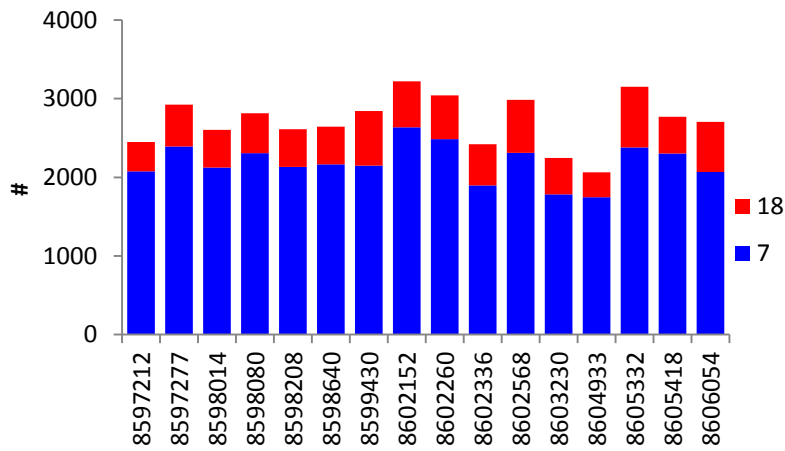

**Figure 7. The number of event hypotheses associated with 7 and 18 stations in cross correlation processing.**

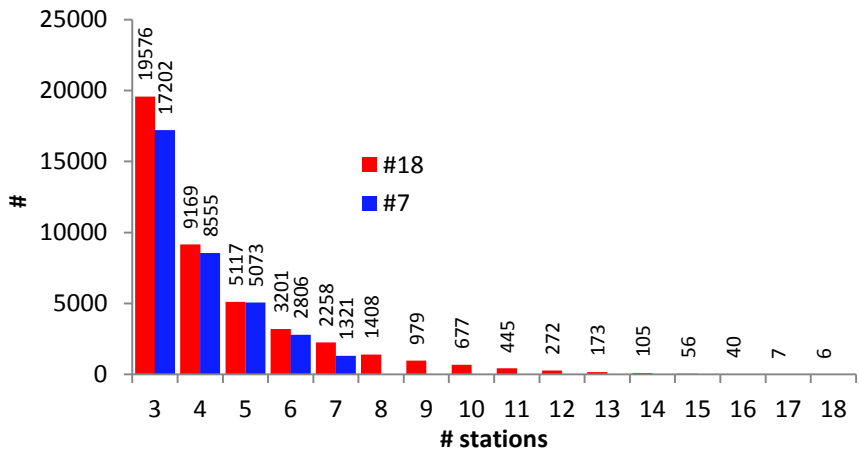

**Figure 8. The number of hypotheses as a function of the number of stations for 7 and 18 stations in cross correlation processing.**

Another important issue is the density of master events. We have selected only 16 masters to cover the area of more than 250,000 km$^2$. Our previous experience shows that the increasing number of masters do improve the completeness of catalog. Ultimately, one can iteratively use all qualified aftershocks (i.e. those having at least three stations with SNR>5) as master events (Bobrov *et al.*, 2012). This may increase the portion of new events to 70% and even more. When a sparse grid of master events is used, one may expect less than 50% of new events. With the incresing masters' density and iterations always starting from the first day, the need in computer and human resources rises dramatically. The rate of false events may also rise since all detections with larger amplitudes are associated at the previous iterations. The number of masters and the use of iterative procedures have to be defined by resources. We have chosen a simplistic approach and the result might be not optimal.

The decay of aftershock sequences follows Omori's law: the rate of aftershocks, $n(t)$, falls as $k/(c+t)^p$, where $k$ and $c$ are constants, and $p$ is between 0.7 and 1.5. The studied Sumatera sequence also follows this law as Figure 9 demosartes. For the REB aftersocks, $p=1.2$ and for the aXSEL $p=1.1$ ($c=0$). Both values are well within the typical range. These curves show that only four to five days after the main shock are the challenge for the *LA*. When the rate of events is less than, say, four per hour, local association does not need to resolve conflicts between hypotheses with the same origin time and based on the same master event. Such conflicts is the highest challenge not only for cross correlation but also for the current IDC pipeline. Globally, almost all standalone events do not interfere in terms of arrivals at the same stations (i.e. doublets are rear).

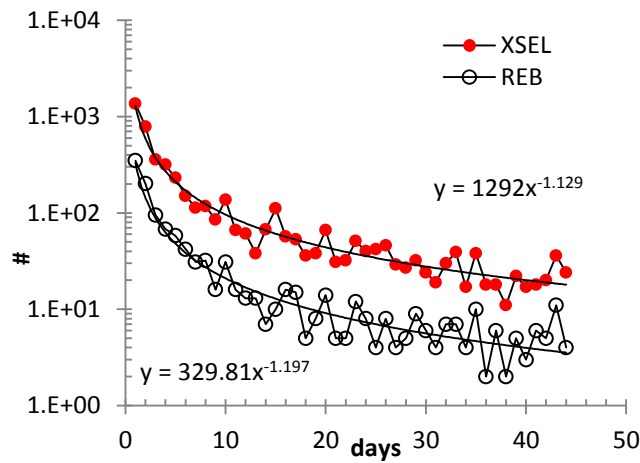

**Figure 9. Omori's law for the REB and automatic XSEL.**

The conflict resolution between similar hypotheses from different masters included two iterations. During the first iteration, we retained all phases which were not associated with the aXSEL events before the conflict resolution. The assumption was that there could be two aftershocks close in time and space but different in size. After the first iteration, there were 3010 hypotheses in the aXSEL. The second iteration uses the same LA procedure and includes only these retained phases. After two iterations, the aXSEL includes 4924 hypotheses, i.e. 1914 events were added to the aXSEL during the second iteration. Overall, only 14% of original hypotheses survive after the conflict resolution. Accordingly, we have two sets of associated detections: those associated with one of 34957 events build by all masters and a smaller subset of detections associated with the aXSEL. There are no human resources to review all 4900+ hypotheses in the aXSEL and we use this automatic bulletin to estimate various statistic properties of the event defining parameters. The obtained estimates may slightly differ from those associated with the subset of valid REB events in the aXSEL. Figure 10 depicts the distribution of three types of hypotheses over master events. (Table 1 lists the number of zero-level hypotehses).

The aXSEL can be split is two parts: the events close in time and space to one or several aftershocks in the REB, and those events which are far from any aftershock in the REB. The latter events are called "new" and define the superiority of the aXSEL. We have consider an aXSEL hypotehsis as matching an REB event when three or more respective arrivals are within ±6 s and the difference of body wave magnitudes is less than 0.7. For the aXSEL, the magnitude is the sum of the master event magnitude plus the network aveargae *RM*. When all events matched by similar arrivals from at least one aftershock in the REB were excluded there were 2763 (1739 after the first iteration) new hypotheses left in the aXSEL. Many events in the aXSEL were built by arrivals in P-coda, with intensive water reflections as a prominent source of strong arrivals in the coda. Since these reflections-based events have arrivals close in time to those from valid REB events, they were excluded after the comparison of the REB and aXSEL. Therefore, the total number of new hypothesis not matched by the REB is not the simple difference between the aXSEL and REB. The distribution of these new events over masters is also shown in Figure 10. The interactive review was focused on the new events.

There are REB events not matched by any of aXSEL events. These are small events with a few stations and, at this stage, we consider them as bogus since they do not have phases correlating with any of the master events. This is a suspicious feature because most of similar REB events are correlating with the masters. There are two major causes for these suspicious events. One is related to wrongly associate seismic phases, which are actually coda waves (reflections) or belong to different sources. Such events are likely mislocated by hundreds kilometres as one can observe in Figure 2. Same effect was found in the Chinese aftershock sequence (Bobrov, Kitov, and Zerbo, 2012). It is difficult to fight against this effect only with signal features currently available in IDC processing. As a matter of fact, all event definition criteria are matched for these likely bogus events. Another reason may be associated with low resolution of the current set of master event. It does not cover the areas of the REB events missing from the aXSEL. This issue is easy to resolve by introducing more masters.

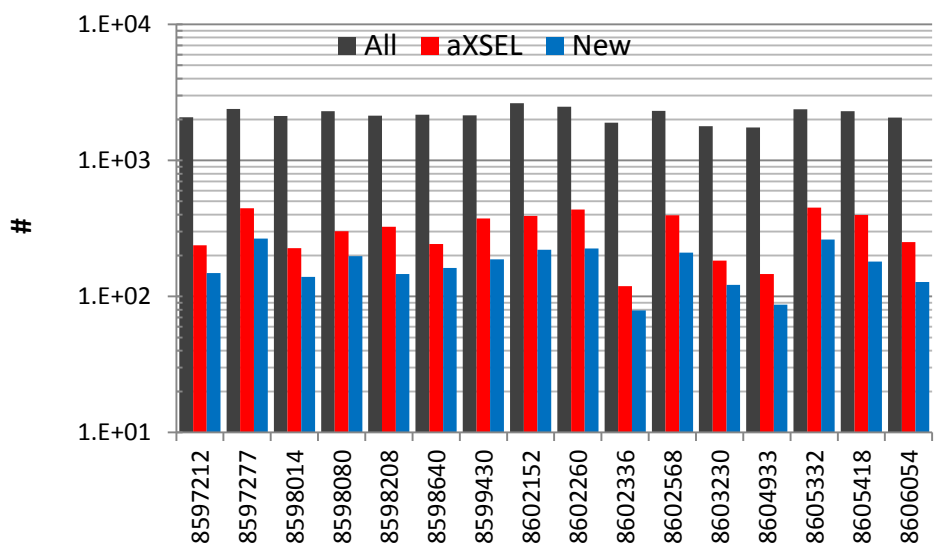

**Figure 10. All events found by all masters, aXSEL and potentially new (extra to the REB) events in the aXSEL. (Notice the logarithm scale.)**

The rate of successful and failed hypotheses is estimated by an "exit poll" approach: from the whole aXSEL we have randomly chosen 10% of events. We expect that the distribution of true and bogus hypotheses in the randomly chosen subset accurately represents the entire population in statistical terms. Then, it is possible to better tune all defining parameters of the cross correlation pipeline. As an aletrnative, we have seleted one master event and reviewed all relevant hypotheses obtained after the first LA iteration. For the monitoring regime, missing the biggest events is of extraordinary significance and we have selected a special subset including all events with 6 and 7 defining stations.

Figure 11 (left panel) comapres the probability density functions (PDF) for the aveage event *CC* (*CC_AVE*), i.e. the sum of individual absolute values of cross correlation coefficient, |*CC*|, divided by the number of stations in the event, as obtained for all events built by the *LA* and those in the automatic XSEL. A larger portion of events in the aXSEL have lower average *CC*s, which peak at 0.35. Obviously, larger events can be built by many master events and thus their input is biased up in relative terms. When only physically unique events are considered, smaller events prevail. There is a linear segment in the aXSEL curve between 0.35 and 0.55, which corresponds to exponential distribution (notice the lin-log scale). This might manifest the exponential dependence of *CC* on the distance between master and slave. One can not exclude that there are real events with the average *CC* of 0.2 to 0.3, but such hypotheses have to be scrutinized. A more reliable estimate of the *CC_AVE* threshold is 0.35. Below 0.35, one should tune the decision line balancing the flux of missed and bogus events. Station-dependent *CC* thresholds have also be involved in the estimation of decision line.

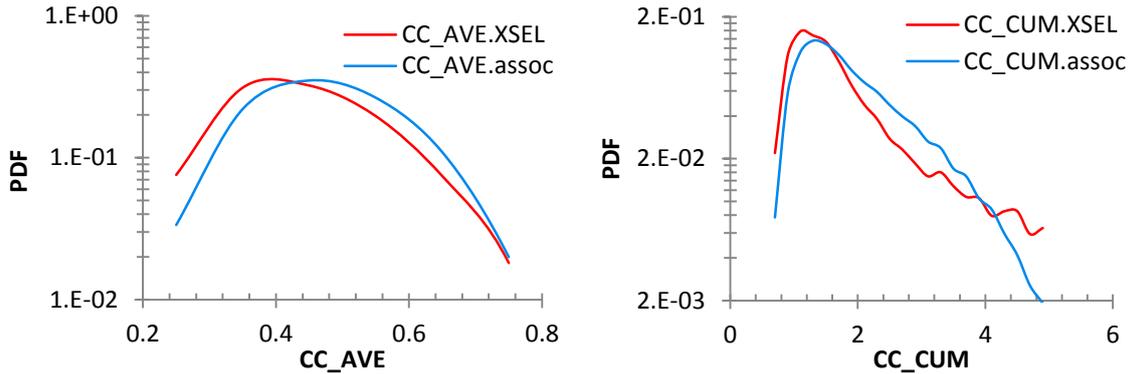

**Figure 11. Left panel: Probability density functions of the avegare *CC* for all associated phases, CC_AVE.assoc, and those in the aXSEL, CC_AVE.XSEL. Right panel: Probability density functions of the cumulative correlation coefficients as obtained from the aXSEL and from all detections associated with REB-ready events. Autocorelation is excluded.**

The cumulative *CC* (*CC_CUM*), the sum of individual absolute values of cross correlation coefficient, in Figure 11 (right panel) expresses the joint input of the defining stations. The PDF for all associated arrivals (blue line) shows a faster fall than that obtained from the automatic XSEL (red line). Both lines have broader intervals of exponential distribution above their respective corner values. At the same time, the corner value of the latter distribution is smaller, i.e. the aXSEL contains a larger portion of smaller and larger events and is below the blue line between 1.2 and 3.2. In this range, the input of misassociated arrivals with small *CC* in the whole list may cause the higher density. The events with such misassociated arrivals eventually miss the aXSEL since the relevant hypotheses are rejected during the conflict resolution between the master events.

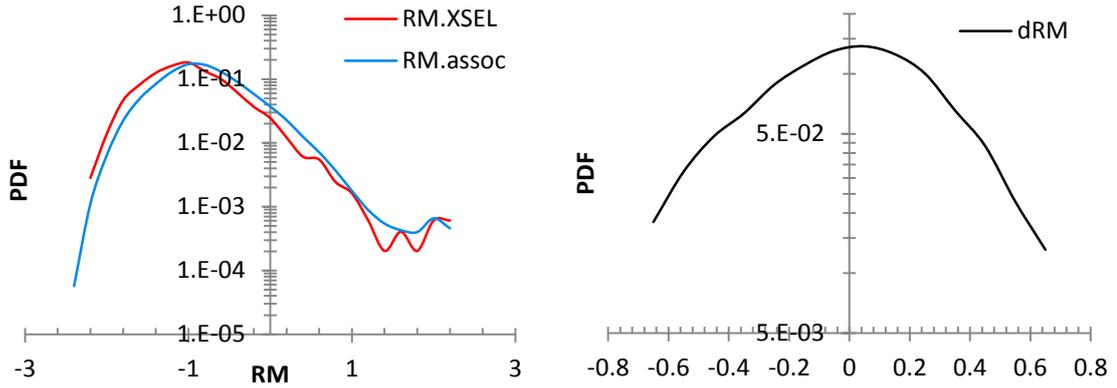

Figure 12. Left panel: Probability density functions of the average relative magnitude for all associated phases, RM.assoc, and those in the aXSEL, RM.XSEL. Right panel: Probability density function of the relative magnitude residual for the automatic XSEL.

The crucial importance of the relative magnitude for event building and magnitude estimation is illustrated in Figures 12 and 13. The relative magnitude averaged over all stations for a given event peaks at -1.2 for the automatic XSEL events (RM.XSEL) and at -1.0 for all events built with 16 masters (RM.assoc). For the smaller average *RM*, both distributions fall very fast and there are no events with *RM* < -2.5. For the larger *RM*, the rate of fall is exponential, as expected from the frequency distribution of the number of larger events. Therefore, one can expect a complete catalogue for all average *RM* estimates above -1.0. For the average magnitude of the master events 4.7, the completeness threshold is 3.7. This is equivalent to the threshold in Figure 5.

The scattering of *RM* station residuals, *dRM*, is slightly skewed, as the right panel of Figure 12 shows. The density of positive residuals falls faster than that of negative ones. The overall distribution evidences that our assumption on the low dispersion of station *RM* residuals (within 0.7) was right and more than 90% lie inside ±0.4. This may put the station *RM* threshold even lower. Figure 13 (left panel) depicts *dRM* scattering for individual stations. The frequency distributions are shown, which may better express the difference in the total number of arrivals. The station behaviour varies is a wide range, with SONM having just a few large positive deviations compensated by a bigger number of negative ones between -0.6 and -0.7. On the contrary, MKAR has many positive deviations around 0.6 and the smallest number of *dRM* estimates below -0.4, which may be associated with an increased noise input biasing the estimates up. Overall, the distributions in Figure 13 allow estimating station-dependent and non-symmetric thresholds for *dRM*. Since we use the automatic XSEL instead of the final one obtained after interactive analysis, all distributions for the REB-ready events may slightly change. We do not expect any major shifts, however.

The right panel of Figure 13 displays station-dependent distributions of *CC* as obtained from the automatic XSEL. This parameter controls the flux and quality of arrivals at all stations. For *CC*, we observe the highest diversity of distributions. All distributions seem to be symmetric in shape and number. Several stations have peaks between 0.4 (CMAR) and 0.6 (MKAR). For MKAR, the peak at the largest |*CC*| is related to its highest SNR for the main shock: when an aXSEL event is built, all associated stations generally have lower *CC*s than that for MKAR. Therefore, for the events in the automatic XSEL, MKAR cannot have low *CC* values, which are present in the pool of all detections. GERES and WRA have weak peaks at 0.6 and 0.4, respectively, and then the distributions are practically constant down to 0.2. ASAR has no peak above 0.2 and its distributions just falls with the absolute *CC* value. Having these distributions one can define individual *CC* thresholds for each of the seven stations.

Figure 14 demonstrates the possibility of another filter to reject inappropriate detections. It shows station-dependent frequency distributions for *CC_SNR* as obtained from the automatic XSEL. Interestingly, all distributions have exponential character above SNR=4.0. This observation implies that the threshold of 4.0 for *CC_SNR* might guarantee that the signal is always good for the aXSEL.

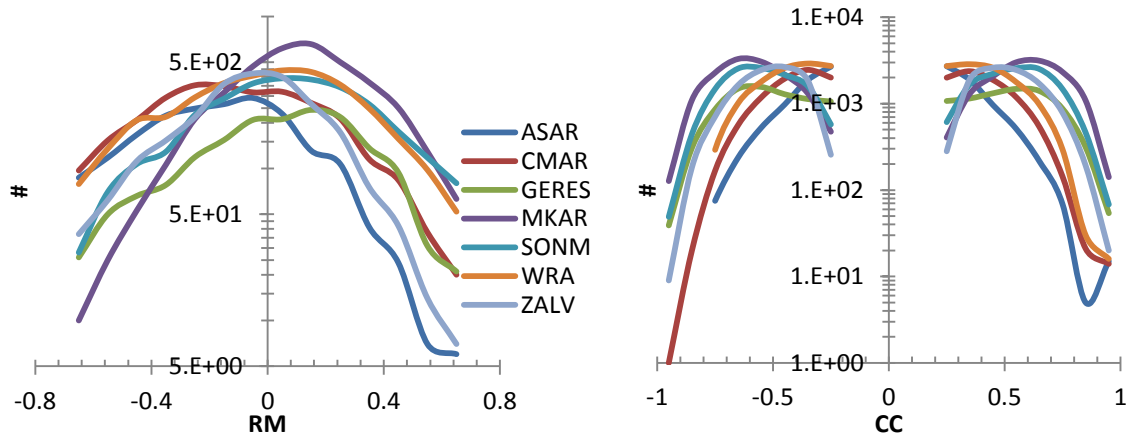

**Figure 13. Left panel: Frequency distribution of the relative magnitude residuals for seven stations as obtained from the automatic XSEL. Right panel: Probability density function of the frequency distribution of *CC* for seven stations as obtained from the automatic XSEL.**

In this study, we apply no of the estimated thresholds and leave this problem for the further investigation at regional and global levels. These thresholds have to be balanced over larger areas in order to equalize the overall monitoring resolution. The CTBT requires homogeneous monitoring threshold. This does not preclude any Member States to retrieve all information from IMS data using any technique including waveform cross correlation.

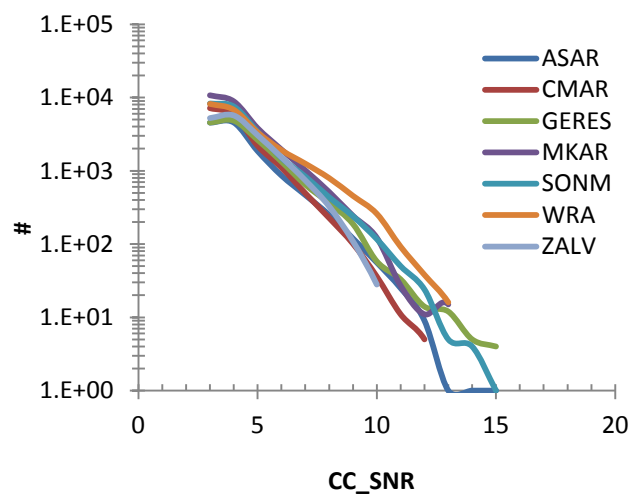

**Figure 14. Frequency distribution of *CC_SNR* for seven stations as obtained from the automatic XSEL.**

After the interactive review, the aXSEL hypotheses are split into two categories. There are hypotheses which have been converted into valid REB events, i.e. the events matching all IDC event definition criteria. Since we used relatively low thresholds of *CC* and *SNR_CC*, there was a larger portion of hypotehses in the aXSEL which failed and no REB events were created. This does not mean that all these hypoteshes are wrong in seimological terms, however. Analysts have to check each associated arrival according to IDC rules and quidlines. These rules prescribe all asociated signals to be visible on the best beams (for array stations) and subjective judgements are allowed. Analsysts approve or reject arrivals according to their understanding and experience on appropriate signal features.

It should be noticed that the best beam, where the maximum signal enhancement is expected, may suffer substantial beam losses due to small deviations of theoretical time delays between individual channels from actual time delays. The cross correlation uses the actual delays and has no losses of this type. This is especially important for weak signals, where the advantage of cross correaltion over beam forming provides more valid signals. Some of these signals are not visible because of beam forming failure. Accordingly, the hypotheses with valid detections obtained by cross correlation but not approved by analysts may fail. But the relevant events do exist. (As an alternative, one may reconsider the EDC.) In this study, we call all failed aXSEL hypotheses "bogus", but only in the narrow sense of IDC rules.

Due to the extremely large workload, the interactive review is not possible for all (~5000) aXSEL events. By construction, the REB-ready events are easier to review. All aXSEL hypotheses include at least three primary stations with arrivals matching the relevant azimuth and slowness uncertainty bounds. Therefore, the analyst has to confirm the existence (visibility) of the arrivals and relocate the events in order to fit travel time residuals. When the number of approved arrivals is less than three, no additional search for arivals at different stations is needed since all hypotheses are already based on the most sensitive stations. The probability to find another station to corroborate an event with the number of defining stations less than three is negligible. (We do not consider here regional 3-C stations which may help in event building but in limited areas.) These are two major problems with the review of SEL3 events, where the search for three primary stations and checking a huge number of alternative solutions for smallest events consume half (if not more) of working time. In any case, the workload related to this study exceeds the limits of available human resources.

As discussed above, we have carried out two different estimates of the true rates of valid (REB) and bogus events in the aXSEL. Both are based on interactive review of a portion of the whole set or "exit poll" statistics. We have selected two subsets of data from the aXSEL. At first, we reviewed all hypotheses resulted from the first LA iteration for one master event (orid=8598208). There are 109 aXSEL events: 72 with 3 defining stations, 22 with 4 stations, 12 with 5 stations, and 3 events with 6 defining stations. This master built no 7-station events.

The other subset includes 250 events randomly chosen from the set of 2763 aXSEL events (~10%) having less than three common (close in time and *RM*) phases with at least one REB event. We imply that these events are not in the REB and thus are potentially new REB events. The possibility to have one or two common phases is not excluded as well as the use of arrivals in coda of strong aftershocks already in the REB. One cannot exclude another independent aftershock with one of a few phases close to those from 1181 REB aftershocks. This situation needs analysts' review.

The randomly chosen hypotheses do not include any of 40 (from 2763) events with six and seven of defining stations. These 40 events are all reviewed separately because of their special importance for nuclear test monitoring – no larger events should be missed. Therefore, the second subset contains two parts and 290 events to review. Figure 15 depicts the distributions of all selected events over the number of defining stations and masters.

These distributions should accurately represent the whole aXSEL set in terms of statistics. Therefore, we also plot the overall statistics in Figure 15, excluding 40 big events.

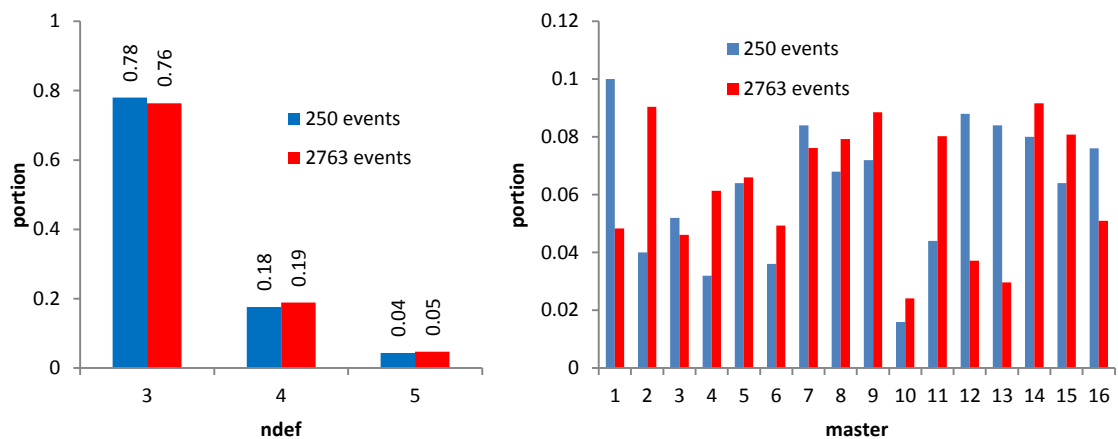

**Figure 15.** Probability density functions of the distribution of event hypotheses over the number of defining stations and master events for the whole set of 2763 events and the randomly chosen subset of 250 events. The overall similarity is retained with the distribution over masters subject to larger variations.

After a standard interactive review of 109 events, there were built 38 new events matching the EDC, i.e. the events missed by the REB. For 17 from 109 events we have found REB counterparts, which were missed by automatic comparison of the REB and aXSEL. These events should be excluded from the set of 109. We cannot define any of them as a new or bogus one because they represent valid REB events. The definition of closeness between aXSEL and REB events might be revised in order to eliminate these 17 events from the aXSEL before the interactive review, but this is not the goal of this study. For the whole set, the success rate is (38/109=) 0.35. Assuming a homogeneous distribution of aftershocks over space and similar performance of the other 15 masters, one can estimate the number of missed REB events in the sequence as ~600. This makes ~50% of the relevant REB. Considering the inhomogeneity of aftershock spatial distribution (the selected master is located on the periphery of the aftershock area) and the imperfectness of the master event choice and coverage (there are master events with much higher SNRs at all seven stations), the number of missed events may rise and is likely closer to the number of REB events.

When 17 events having valid REB counterparts are subtracted, the rate of new events is 38/92=0.41, i.e. approximately 60% of the aXSEL events are bogus. Fortunately, there are quantitative parameters, which allow screening out most of bogus events without any significant loss in the success rate. It is highly important that the events we review are small (with a few exceptions) compared to the events in the REB, and thus, are most challenging for nuclear test monitoring. It should be also mentioned that the reviewed aXSEL events are most difficult for interactive analysis since they are characterized by weaker signals and smaller number of stations. The aXSEL hypotheses associated with the REB aftershocks have from four to seven (from seven) defining stations with clear signals. These hypotheses are easier to review interactively.

The set of 290 hypotheses was also reviewed interactively. We have built 1 new event from 7 hypotheses with seven defining stations and 10 from 31 with six stations. Seven events from these 40 had weak signals and can be treated as bogus. Twenty two events had REB counterparts. The waveforms from two big events with six primary array stations, which were found several days after the main shock, are shown in Figure 16. The corresponding signals are clear and reliable – such events should be missed in no case. Together with five 7-station events found during the first day, one has 16 (!) big events missed by IDC automatic processing. We do not estimate the rate of success for these bigger aXSEL events: no one big event should be missed.

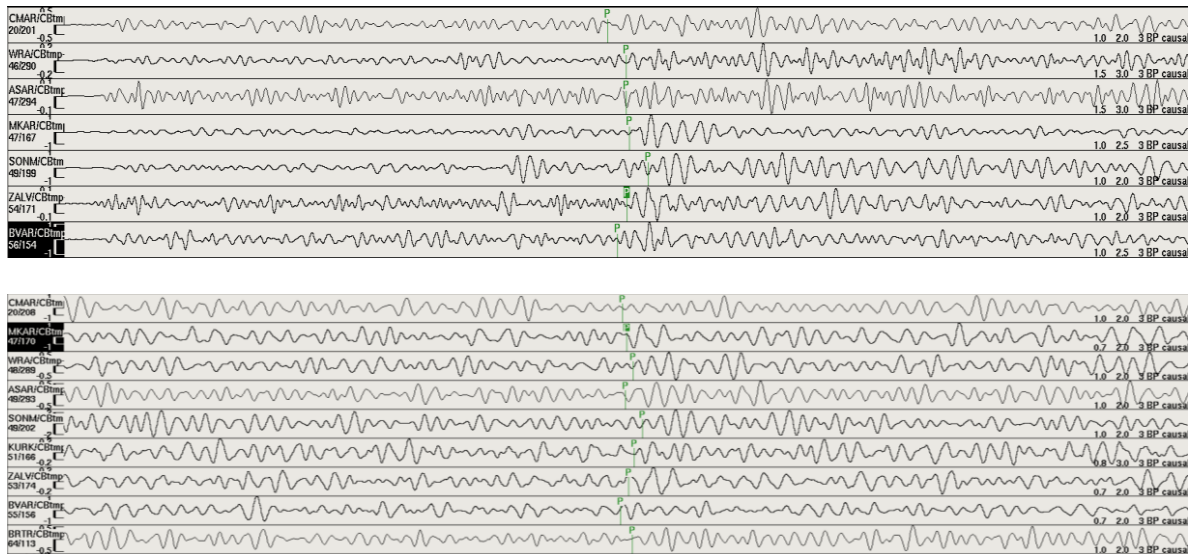

**Figure 16. Two new REB events with six defining stations in the aXSEL, which were found three and six days after the main shock, respectively.**

There were 70 new REB-ready events built from the set of 250 hypotheses, with only 88 bogus ones. This makes 92 hypotheses to be found in coda of big aftershocks or associated with REB events. The succes rate for the whole set is (70/250=) 0.28. When scaled to ~2700 hypotheses, the expected number of new events is ~750. This number is higher when that determined from one master event. It may express a better performance of other 15 masters and concentration of aftershocks near the main shock. When 92 events associated with the REB are excluded, the succes rate is 44%, i.e. very close to that from one master.

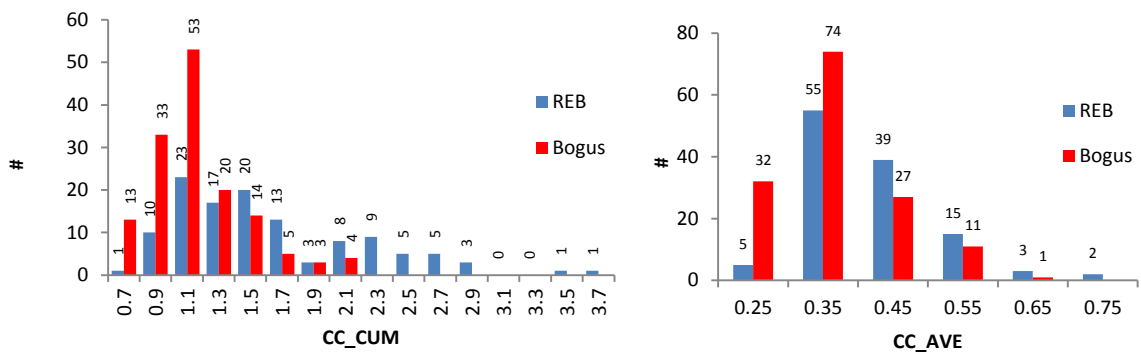

**Figure 17. Frequency distributions over *CC_CUM* (left panel) and *CC_AVE* (right panel) of 119 aXSEL events converted into the new REB events and 145 aXSEL events which can not be converted into REB events (bogus) as obtained by interactive review of 399 selected aXSEL events.**

In total, we have 119 (38+11+70) new REB events and 145 bogus events since 4 hypotheses are common for both studied sets. We intentionally allowed weak signals with small *CC* and *SNR_CC* to be created. As a consequence, there are many bogus (in terms of REB) events built in the aXSEL. Figure 17 depicts the distribution of new and bogus events over the cumulative CC, *CC_CUM*, and average CC, *CC_AVE*. The new and bogus events distribution both peak between *CC_CUM* 1.0 and 1.2, but for the bogus distribution the peak is prominent. Overall, only 34 new events (29%) and 99 bogus events (68%) have *CC_CUM* below 1.2. This is an effective criterion to screen most bogus events out before interactive review, and without a big loss of valid hypotheses. For *CC_AVE*, both distributions are similar with 32 bogus events between 0.2 and 0.3 to be screened out with the loss of 5 new REB events.

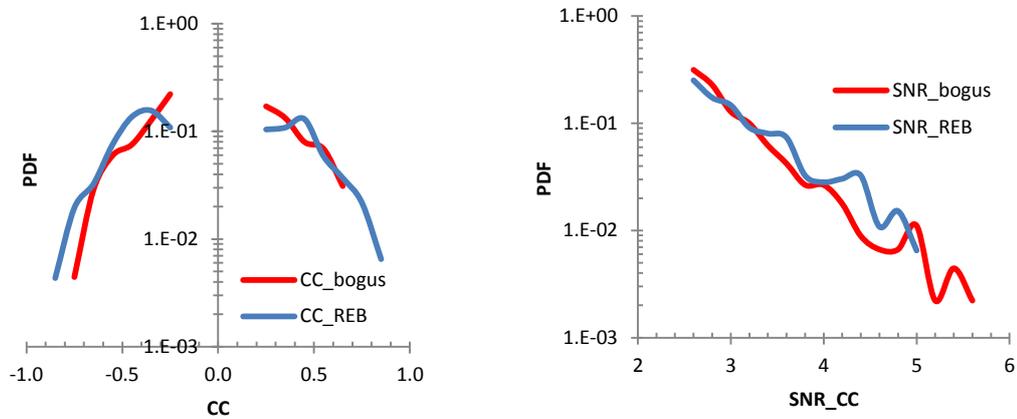

**Figure 18. Probability density functions for *CC* (left panel) and *SNR_CC* (right panel) as obtained from defining phases at seven arrays stations for the new REB events and the events which do not meet some REB criteria (bogus).**

A more efficient way to eliminate bogus events from the aXSEL might be based on the distribution on individual *CC*s and *SNR_CC*s. Figure 18 displays the probability density functions as obtained from the signals associated with the new REB (461 arrivals) and bogus (451 arrivals) events. Interstingly, the negative and positive *CC* branches are not symmetric and there are no bogus events with *CC* above 0.65. Overall, the bogus events include more low-CC and less high-SNR arrivals. It is time to stress again that the bogus events are likely to be seismilogically sound and just do not fit the EDC designed for the the Global Association which allows much larger uncertainty bounds than those for the LA.

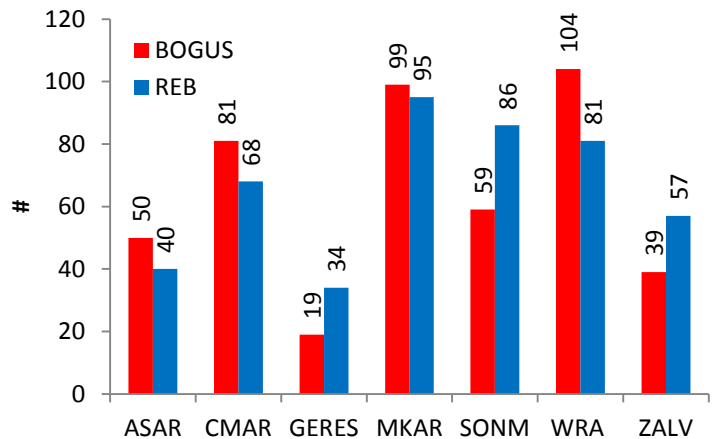

**Figure 19. The number of arrivals at seven array stations as associaited with the aXSEL events, which were converted into the new REB events, and those aXSEL events, which can not be converted (not meeting some of the EDC).**

Figure 19 shows that there are three most effective stations: MKAR, SONM, and WRA. Three stations (GERES, SONM, and ZALV) have higher numbers of arrivals associated with the new REB events than with the bogus events. The largest number of detections associated with the bogus events comes from WRA and MKAR. These two stations are most sensitive and often create hypotheses with weak third signals from any other station. Such hypotheses have a higher probability to fail. For screening purposes, one can use the absolute probability of a detection at a given station and with given *CC* and *SNR_CC* to be associated with an REB event as well as the conditional probability distributions of various station configurations for the REB and bogus events. The assessment of all probability distributions has to be carried out using all aXSEL events converted into valid REB events, however. We are going to obtain such estimates at the next stage of cross correlation research, when the global REB is assessed by machine learning tools.

After the interactive review, we have 119 new REB events. Figure 20 plots their locations. There are several mislocated events far away from the zone of aftershocks. The IDC location of weak events with only three stations is not reliable. Similar outliers were reported in the REB. All mislocated events have (likely better) cross correlation locations as associated with 16 master events and the circular grid of 18 stations around each of them.

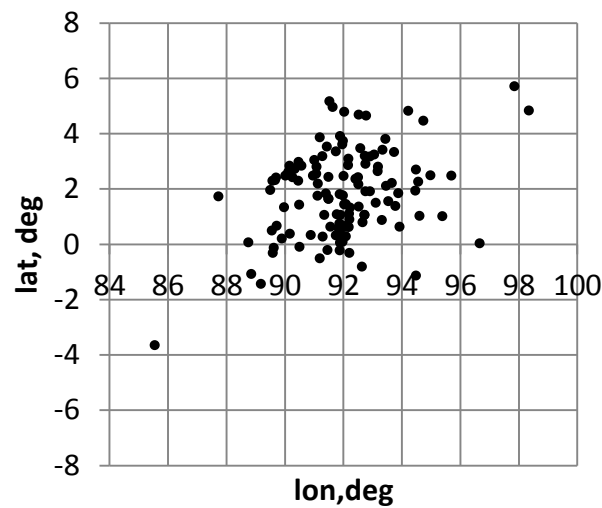

**Figure 20. Locations of 119 new REB events as estimated by standard IDC location program.**

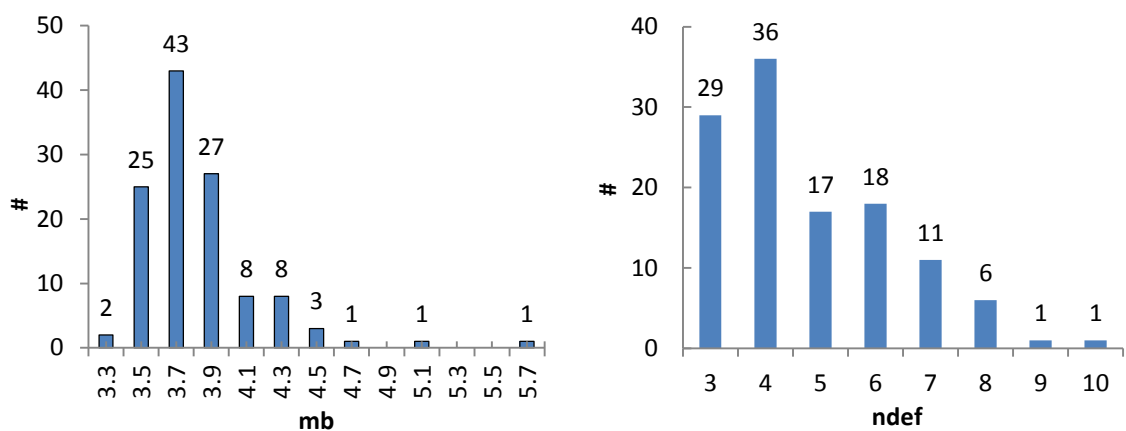

**Figure 21. Frequency distributions over $m_b$(IDC) (left panel) and ndef (right panel) of 119 new REB events.**

Figure 21 depicts two distributions obtained from the new REB events: over $m_b$(IDC) and the number of defining stations (from the seven used for cross correlation). The former distribution peaks between 3.6 and 3.8. This is an expected result, as Figure 5 suggests. The aXSEL is likely complete to magnitude ~3.8. The new REB events reproduce this feature of the aXSEL and improve the REB completeness to the same level. When extrapolated to the whole aXSEL, the number of events with $m_b$(IDC) 3.6 to 3.8 is between 250 and 300, i.e. about a half of the deficit observed in the REB curve in this magnitude bin. The bin between 3.8 and 4.0 is almost fully complete. Therefore, the cross correlation technique reduces the monitoring threshold. Moreover, it does not miss bigger events: there are several events with magnitude above 4.2 found by cross correlation.

Several bigger events have more than 7 associated IMS stations, some of them are 3-C auxiliary stations at regional distances. We exclude all arrivals beyond the seven involved stations and estimated the distribution of arrivals over SNR. This is standard IDC SNR as obtained from original waveforms using the origin beam (Coyne *et al.,* 2012). The IDC distribution peaks at 2.2, which is less than the smallest *SNR_CC* (2.5). Many arrivals in Figure 22 have SNR below 2.0. We have already discussed the beam loss as the cause of poor performace of the beam forming. However, there are IDC detections with SNR higher than 3. This observation evidences that IDC automatic detector misses many signals with realtively high SNRs. Figure 22 proves that cross correlation is a superior detection tool. For weak signals, due to destructive interference of not perfectly shifted individual waveforms during beam forming there is no visible signals. The corresponding cross correlation traces show signals with prominent SNRs. One may use the CC traces for the purposes of interactive review.

**Discussion**

We have created an aXSEL for the aftershock sequence of the April 11, 2012 Sumatera earthquake. The aXSEL includes 4924 event hypotheses. Almost all of 1181 aftershocks from the REB for the same period are matched by one or more events from the aXSEL. There are more than 2750 events hypotheses extra to the REB ones, i.e. there are no REB events close in time and space. Severe resource limits did not allow a full interactive review of these hypotheses according to IDC rules and guidelines. We have applied an "exit poll" technique in order to statistically characterize the rate of valid and bogus events in the aXSEL.

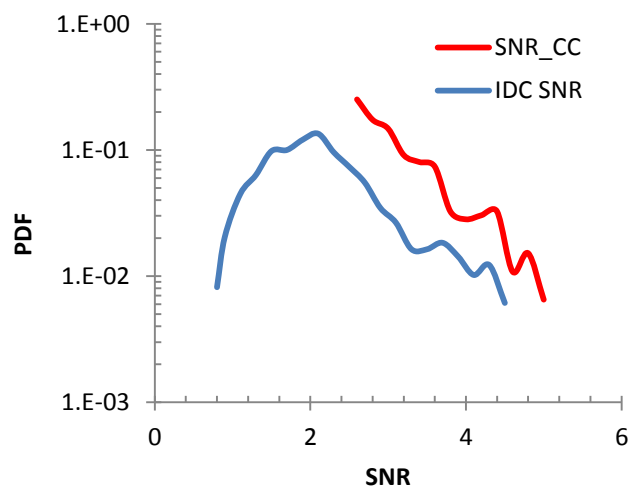

**Figure 22. Probability density functions for SNRs determined by standard IDC processing of original waveforms (IDC SNR) and with cross correlation (SNR_CC) for 119 new events.**

For one master event, we have conducted a comprehensive review. We have found 38 new REB events from 92 hypotheses not associated with any aftershock in the REB. It makes around 600 new events overall. The review of 250 hypotheses randomly chosen from 2763 aXSEL events (~10%) has revealed 71 new REB events. This implies a larger number of events missed by the REB: ~750. Therefore, one can expect between 50% and 65% of missed events for the case of 16 masters. Since the choice of master events was not optimal and we did not apply iterative procedures with increasing number of masters, the true portion of events missed in the REB might reach 100%. Moreover, we have built 16 new REB events with six and more defining stations.

As a part of the exercise, several IDC analysts reported that they were able to easily accept or reject the hypotheses from aXSEL. They were able to quickly check for three or more clear arrivals, confirmed by the *f-k* attributes. Analysts searched for additional, non-associated, stations but these were found in only a couple of instances. The workload was reduced by a factor of two and even more. Considering the hypotheses reviewed by the analysts as the most difficult ones we expect a significant reduction in the workload associated with the interactive analysis even with the doubled number of events in the REB.

We have found many low magnitude events and likely reduced the monitoring threshold for the CTBT by 0.4 units of magnitude. These small events are most difficult to find automatically and review interactively. In order to better estimate the portion of REB-ready events and the rate of false alarms we are going to select a small subset mimicking the distribution of hypothesis over master events and the number of defining stations (from 3 to 7).

Overall, the waveform cross correlation technique has proved its superior performance relative to automatic processing at the IDC. This technique has found more REB-ready events (XSEL) than the REB included for the same period. The set of new events included five new aftershocks during the first two days after the main shock, which were based on all seven IMS array stations used for cross correlation. These five events have $m_b$(IDC) between 4.3 and 5.0. The XSEL also excludes those REB events, which do not demonstrate a reliable level of cross correlation with any neighbouring events. Therefore, the quality of XSEL is likely higher than that of the REB.

The exercise with a large aftershock sequence is an important step on the way to global coverage. The cross correlation technique is based on master events, which are available only in seismic regions. In practical terms, seismicity is a stationary process with repeating events within well-constrained areas. This makes the choice of masters easy for 99% of events in the REB. We have already selected a set of master events in the North Atlantic after calculation of cross correlation coefficients for all REB events after 2009 (Bobrov *et al.*, 2012). This is a working prototype for the global coverage. The procedure has been tuned and tested to optimise the performance of automatic processing for the purposes of interactive review.

The areas without master events represent a challenge for cross correlation. Fortunately, the method is powerful enough to provide a better monitoring capability even for aseismic areas. There are three possibilities based on the dependence of cross correlation coefficient on signal shape and time delays between individual channels of an array.

One may use any master event for relatively large area (say, 500 km in radius) just changing theoretical time delays in the master template and studied waveform. This procedure preserves the empirical time shifts between channels, which are the most important advantage of cross correlation. In this regard, we are going to replace all 16 master events for the Sumatera aftershock sequence with one best master (orid=8602152). Instead of unevenly distributed empirical masters, we will introduce a uniform rectangular grid with 70 km to 100 km spacing where all waveforms at individual channels are the same but shifted by theoretical times for given grid points and stations. Since the continuous waveforms are shifted by the same theoretical times the calculation of cross correlation coefficient in frequency domain will be reduced just to one master event. Overall, the use of empirical master events beyond their actual positions allows substantial reduction in the need for computer resources and extends the zone covered by cross correlation by hundreds kilometres.

There are extensive (also continental) aseismic areas where one cannot extrapolate the influence of empirical master events. Then two approaches are feasible. One is based on synthetic seismograms for master/stations pairs. Before it can be used for cross correlation, the accuracy of this method should be tested with actual waveforms in seismic areas. The overall experience with synthetic seismograms is encouraging. Since the most important for cross correlation at array stations are the relative time delays between individual sensors, which are associated with local velocity structure beneath seismic stations, one can use local seismic events for the delays' calibration. Then the empirical time shifts could be scaled to any distance by synthetic seismograms.

The second approach was proposed by Harris and Paik (2006). The entire diversity of signal shapes for a given magnitude range can be decomposed into a few orthonormal functions similar to the decomposition into a Fourier series. They also proposed the algorithm for easy implementation. Therefore, one can use the set of orthonormal functions instead of master events and use the sum of coefficients or the maximum coefficient of the decomposition to best represent the sought signal. Cross correlation coefficients can be calculated accordingly. This procedure should also be tested using actual master events.

The location accuracy is one of the most important characteristics of REB events. This is associated with the area of on-site inspections limited by 1000 km$^2$, as defined by the CTBT. The use of regular master event grid and the resolution of conflicts between adjacent masters allows very accurate location even (and especially) for the smallest events. Moreover, confidence ellipses of the corresponding master events are also the ellipses for the slaves. Regional cross correlation studies documented the location improvement by an order of magnitude and event more for small events. We expect the same effect globally, when the cross correlation pipeline becomes operational.

In terms of computer resources, to produce the aXSEL for the studied aftershock sequence one needed only one regular CPU, which can run as many as twenty master events with templates at seven stations in real time. The process of master events selection can be easily optimized and automated to the level of superficial review only. The progress in computer technology (e.g. GPU), the optimization of master event distribution, and the use of synthetic and replicated templates by neighbouring masters may reduce the need in computer resources by orders of magnitude.

The rate of valid and bogus events should be optimized globally. For each master/station pair we have to define a large number of thresholds: time window, frequency bands, *CC*, *CC_SNR*, azimuth and slowness residual, *F*-statistics, *RM*, the distance between sequential detections with similar characteristics. For the *LA*, the event building procedure (i.e. the tightness in time and *RM*) should be tuned to meet the requirement of the most intensive seismicity after catastrophic earthquakes. We have also to define the minimum time between collocated events. The conflict resolution between similar hypotheses obtained by different masters is currently a challenge. One may try to split larger events with many masters (larger average *RM* as well) and smaller events with one or a few masters. It might be possible to resolve all issues using the machine learning, when having enough data.

We did no touch upon a number of problems associated with regional stations and phases. Regional velocity structure is characterized by large amplitude variations, which reduce waveform similarity and thus cross correlation to several kilometres. Since the IMS seismic network is a regional one with three stations within 20$^o$ from any place within continents, the input of 3-C stations and auxiliary arrays cannot be neglected. Fortunately, the experience with regional stations and networks is extensive and may be borrowed and implemented directly under the IDC framework.

All these advantages of the cross correlation technique and the absence of significant problems in automatic and interactive processing urge the replacement of the current processing pipeline. As a start point, all aftershock sequences at the IDC could be processed by the cross correlation technique.

**Acknowledgements**

The authors are grateful to all analysts at the IDC who have built the REB and a number of XSEL events.